\documentclass[preprints,article,accept,LaTeX,dvi2pdf,moreauthors]{Definitions/mdpi} 
\firstpage{1} 
\pubvolume{1}
\issuenum{1}
\articlenumber{0}
\pubyear{2022}
\copyrightyear{2022}
\datereceived{} 
\dateaccepted{} 
\datepublished{} 
\hreflink{https://doi.org/10.3390/\linebreak
universe8010052} 

\usepackage{amssymb}	
\usepackage{amsfonts}
\usepackage{palatino}

\Title{Estimating the Cosmological Constant from Shadows of Kerr--de Sitter Black Holes}

\TitleCitation{Estimating the Cosmological Constant from Shadows of Kerr--de Sitter Black Holes}

\Author{{Misba} Afrin $^{1,}$*$^{,\dagger}$\href{https://orcid.org/0000-0001-5545-3507}{\orcidicon} and Sushant {G.} Ghosh $^{1,2,\dagger}$\href{https://orcid.org/0000-0002-0835-3690}{\orcidicon}}

\AuthorNames{Misba Afrin and Sushant G. Ghosh}

\AuthorCitation{Afrin, M.; Ghosh, S.G.}

\address{%
$^{1}$ \quad Centre for Theoretical Physics,
Jamia Millia Islamia, New Delhi 110025, India; sghosh2@jmi.ac.in\\
$^{2}$ \quad Astrophysics and Cosmology Research Unit, School of Mathematics, Statistics and Computer Science, \\
 \quad\;\; University of KwaZulu-Natal, Private Bag 54001, Durban 4000, South Africa}

\corres{Correspondence: me.misba@gmail.com or misba@ctp-jamia.res.in}

\firstnote{These authors contributed equally to this work.}

\abstract{The Event Horizon Telescope collaboration has revealed the first direct image  of a black hole, as per the shadow of a Kerr black hole of general relativity. However,  other Kerr-like rotating black holes of modified gravity theories cannot be ignored, and they are essential as they offer an arena in which these theories can be tested through astrophysical observation. This motivates us to investigate asymptotically de Sitter rotating black holes wherein interpreting the cosmological constant $\Lambda$ as the vacuum energy leads to a deformation in the vicinity of a black hole---new Kerr--de Sitter solution, which has a richer geometric structure than the original one.  We derive an analytical formula necessary for the shadow of the new Kerr--de Sitter black holes and then visualize the shadow of black holes for various parameters for an observer at given coordinates $(r_0, \theta_0)$ in the domain $(r_0,\; r_c)$ and estimate the cosmological constant $\Lambda$  from its shadow observables. The shadow observables of the new Kerr--de Sitter black holes significantly deviate from the corresponding observables of the Kerr--de Sitter black hole over an appreciable range of the parameter space. Interestingly, we find a finite parameter space for ($\Lambda$, $a$) where the observables of the two black holes are indistinguishable.}


\begin{document}
\section{Introduction}\label{Intro}
Because of the black hole's defining property at the event horizon and the surrounding photon region, it casts a dark region over the observer's celestial sky, which is known as a shadow \citep{Bardeen:1973tla,Falcke:1999pj}, which is an optical appearance cast by a black hole when it is in front of a distant luminous source.  It is a natural result of Einstein's theory of general relativity (GR), so it can provide us with information on the  properties of the black hole. The seminal works of Synge \citep{Synge:1966okc} and Luminet \citep{Luminet:1979nyg} led the foundation for the shadows of black holes.  Synge \citep{Synge:1966okc} was the first to study the angular radius of the photon capture region around the Schwarzschild black hole. Interestingly, the photon sphere, in the case of a Kerr black hole, becomes a "photon region" which is filled by spherical light-like geodesics at a sphere $r=$ constant.  Bardeen \cite{Bardeen:1973tla} first correctly analysed the shadow of the Kerr black hole and showed that the spin would distort the shape of the shadow, and that the deviation of the shadow, from a circle, is proportional to the spin of the black hole.  Furthermore, Teo \citep{Teo:2020sey} analysed pictures of individual spherical light-like geodesics in the Kerr spacetime.
The application of shadow in unravelling gravity's useful near-horizon features is a valuable tool for testing GR. Thereby, the discussion was extended for other black holes, e.g., for the Kerr--Newman spacetime
\citep{deVries2000}, for $\delta=2$ Tomimatsu--Sato spacetimes
\citep{Bambi:2010hf}, for black holes in extended Chern--Simons
modified gravity \citep{Amarilla:2010zq}, in a Randall--Sundrum
braneworld scenario~\citep{Amarilla:2011fx}, and a Kaluza--Klein
rotating dilaton black hole \citep{Amarilla:2013sj},
for the Kerr--NUT \mbox{spacetime \citep{Abdujabbarov:2012bn}},
for multi-black holes \citep{Yumoto:2012kz},  for regular black holes \citep{Li:2013jra, Kumar:2018ple, Kumar:2020yem, Kumar:2019pjp, Ghosh:2020ece, Amir:2020fpa, Abdujabbarov:2016hnw,Amir:2016cen} and for higher-dimensional black holes~\citep{Papnoi:2014aaa, Ahmed:2020dzj,Ahmed:2020ifa, Amir:2017slq,Eiroa:2017uuq,Vagnozzi:2019apd,Banerjee:2019nnj}.
The shadows of modified theories of gravity black holes cast smaller and more distorted shadows when compared with that of the Kerr black hole \citep{Amarilla:2010zq,Amarilla:2011fx,Amarilla:2013sj,Amir:2017slq,Singh:2017vfr,Mizuno:2018lxz,Ghosh:2020ece,Kumar:2020owy,Afrin:2021imp}.   Moreover,  the spin, mass parameter and possibly other parameters of black holes can be calculated \citep{Hioki:2009na,Tsupko:2017rdo,Cunha:2019dwb,Cunha:2019ikd,Kumar:2018ple,Khodadi:2020jij}. In addition, shadow also finds applications in testing theories of gravity \citep{Kramer:2004hd}, and  Johannsen and Psaltis
\citep{Johannsen:2010ru} explored how to test
the no-hair theorem using the shadow of a black hole. Most investigations have been largely based
on ray tracing in the respective spacetimes, rather than on analytical
studies of the geodesic equation under the assumption that the observer is
at infinity.  This prescription does not work when the black holes are no longer asymptotically flat but instead when it is asymptotically dS$/$AdS, and a cosmological horizon encompassing the event horizon is present \citep{Perlick:2018iye, Kumar:2017tdw, Neves:2020doc,Maluf:2020kgf, Grenzebach:2014fha,Stuchlik:2018qyz,Charbulak:2017bpj}.  Therefore,  the observer is located at the so-called domain of external communication, which is the region defined between the event horizon and the cosmological horizon in asymptotically de Sitter spacetimes \citep{Perlick:2018iye, Kumar:2017tdw}.
Indeed, Rindler and Ishak \citep{Rindler:2007zz} realized that the cosmological constant $\Lambda$ does influence the gravitational lensing features because it changes the angle measurements as well as the angular radius of the shadow  \cite{Perlick:2018iye}.

This paper investigates shadows of recently obtained Kerr black holes with a cosmological constant---new Kerr--de Sitter spacetimes \citep{Ovalle:2021jzf}---and shows our prescription's applicability for the determination of the black hole parameters, emphasizing the comparison with the original Kerr--de Sitter black hole shadows.  We found that the horizons' structure contrast from the original Kerr--de Sitter horizons and expression for the photon sphere radii are also different from those in the original Kerr--de Sitter. Furthermore, we found changes in the photon regions of new Kerr--de Sitter black holes, which affect the black hole shadows.
 \begin{figure}[t]
    \centering
    \includegraphics[scale=0.8]{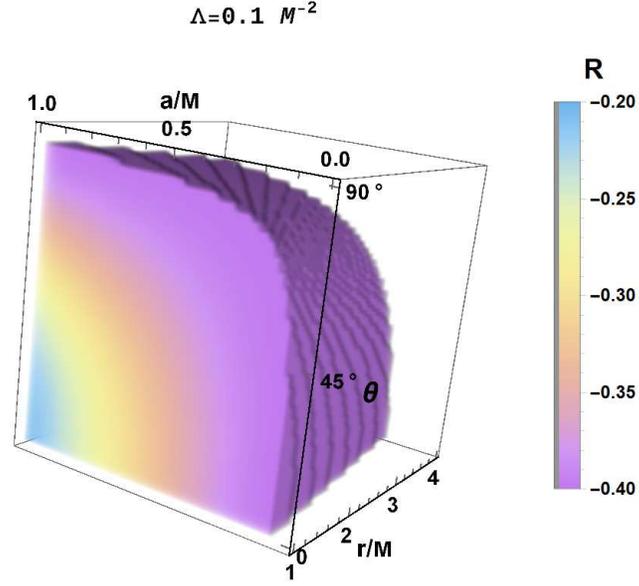}
    \caption{Density plot of the curvature in the vicinity of the Kerr--de Sitter black hole elucidating the warping effect.}
    \label{Curvature}
\end{figure}
\section{Kerr--de Sitter Black Holes}\label{Sec2}
The metric of Kerr--de Sitter in the Boyer--Lindquist coordinates reads \citep{Carter:1973rla}:
\begin{align}\label{metric_ original}
ds^{2}=-&\left[\frac{\Delta_r -a^2 \sin{\theta}^2 \Delta_{\theta}}{\Xi^2\Sigma}\right]dt^{2}+\frac{\Sigma}{\Delta_r}dr^{2}+\frac{\Sigma}{\Delta_{\theta}} d\theta^{2}
-2a\sin^2{\theta}\left[(r^2+a^2)^2\Delta_{\theta}-\Delta_{r} \right]dtd\phi\nonumber\\
 +& \frac{\sin^2{\theta} }{\Xi^2\Sigma}\left[(r^2+a^2)^2\Delta_{\theta}- a^2\sin^2{\theta}\Delta_{r} \right]d\phi^{2},
\end{align}
\begin{figure}
\begin{adjustwidth}{-\extralength}{0cm}
\centering
    \begin{tabular}{c c}
    \includegraphics[scale=0.9]{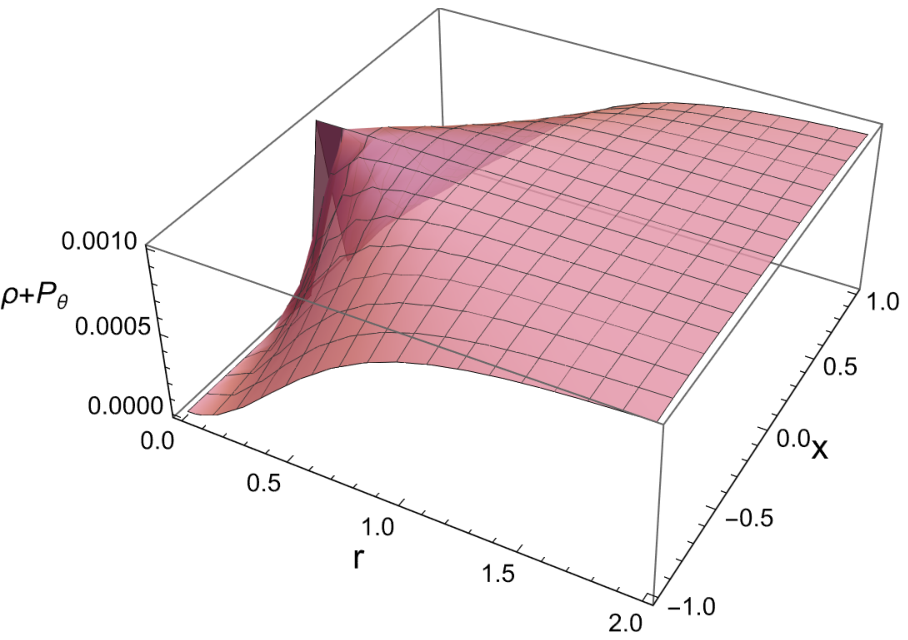}&
	\includegraphics[scale=0.9]{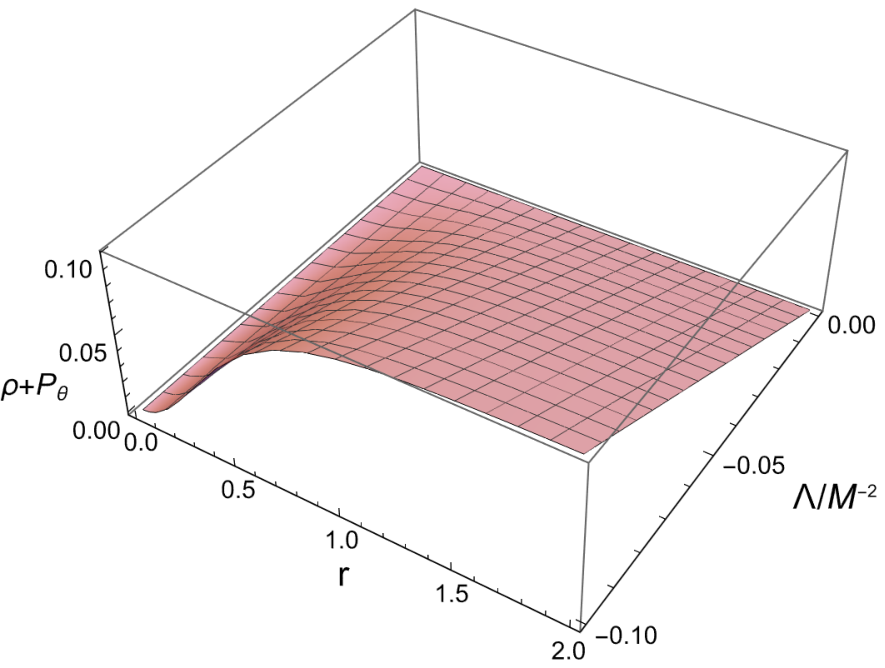}
\end{tabular}
\end{adjustwidth}
\caption{Plots of $\rho+P_\theta= \rho+P_\phi$ with $r$  for different values of $x=\cos{\theta}$ and  $\Lambda/M^{-2}=-0.001$ ({\textbf{left}}) and  for different values of $\Lambda/M^{-2}$ and $\theta=\pi/4$ ({\textbf{right}}). The value of spin parameter in the above plots is $a/M=0.9$.}\label{rho_P_plot}	
\end{figure}
where the various function in  metric (\ref{metric_ original}) are given by, $\Delta_r=r^{2}+a^{2}-2Mr-{[\Lambda r^{2}(r^2+a^2)]}/{3}$, $\Delta_{\theta}=1+{(\Lambda} a^{2}\cos^{2}\theta)/{3}$, $\Sigma=r^2+a^2\cos^2{\theta}$ and $\Xi=1+{(\Lambda a^{2})}/{3}$. The $\Lambda>0$ is positive cosmological constant, $M$ and $a$ are respectively the black hole mass $M$ and spin parameter.
The metric (\ref{metric_ original}) is an exact $\Lambda$-vaccum solution of the Einstein equations: 
\begin{equation}\label{EinsteinEq_original}
    R_{ab}=-\Lambda g_{ab}.
\end{equation}
This was first found by Carter \citep{Carter:1973rla}. In the limit $a\to0$, $M\neq0$, the Kerr--de Sitter results to those of Schwarzschild de Sitter \citep{Bousso:1997wi}, whereas setting further $M=0$ leads to de Sitter metric. However, $M=0$ alone results in a metric diffeomorphic to de Sitter spacetime; thereby the metric (\ref{metric_ original}) is asymptotically de Sitter \citep{Galloway:2007pm}.
\begin{figure}[t]
\begin{adjustwidth}{-\extralength}{0cm}
\centering
    \begin{tabular}{c c}
         \hspace{-10mm}\includegraphics[scale=0.85]{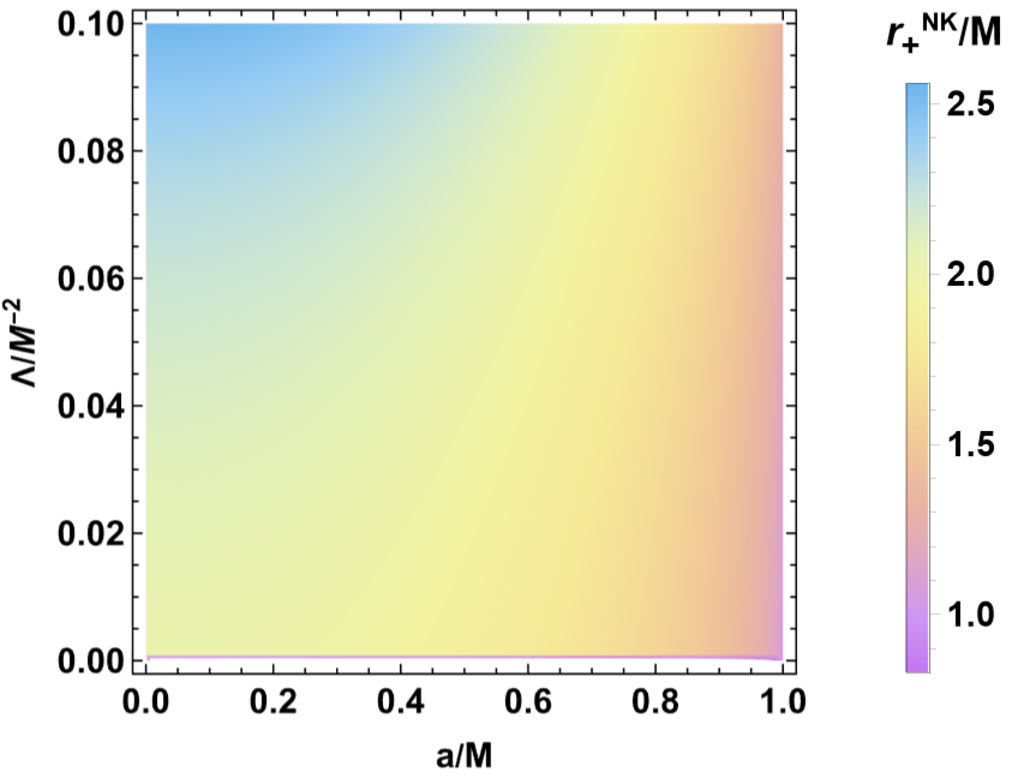}&
         \includegraphics[scale=0.85]{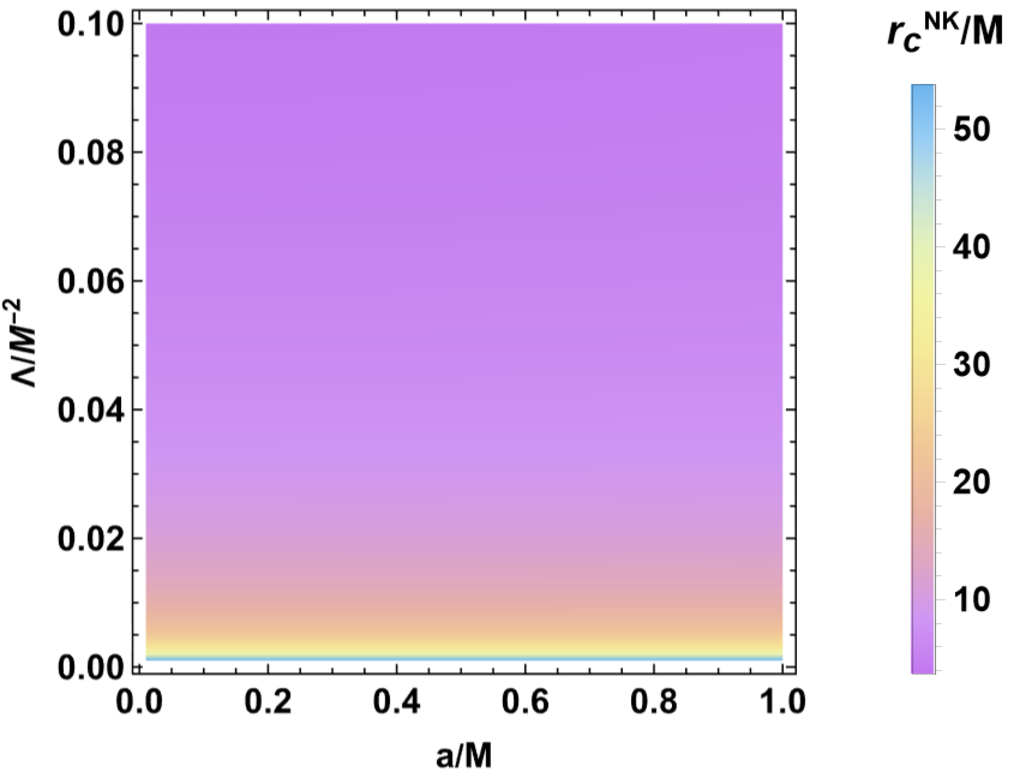}
    \end{tabular}
\end{adjustwidth}
\caption{\label{horizon_plot}Density plot of event horizon ({\textbf{{left}}}) and cosmological horizon ({\textbf{right}}) of the new Kerr--de Sitter black holes shown in parameter space. For a given $a$ and $\Lambda$, we can uniquely determine the horizons.}
\end{figure}
\begin{figure}[t]
\begin{adjustwidth}{-\extralength}{0cm}
\centering
    \begin{tabular}{c c}
         \hspace{-10mm}\includegraphics[scale=0.85]{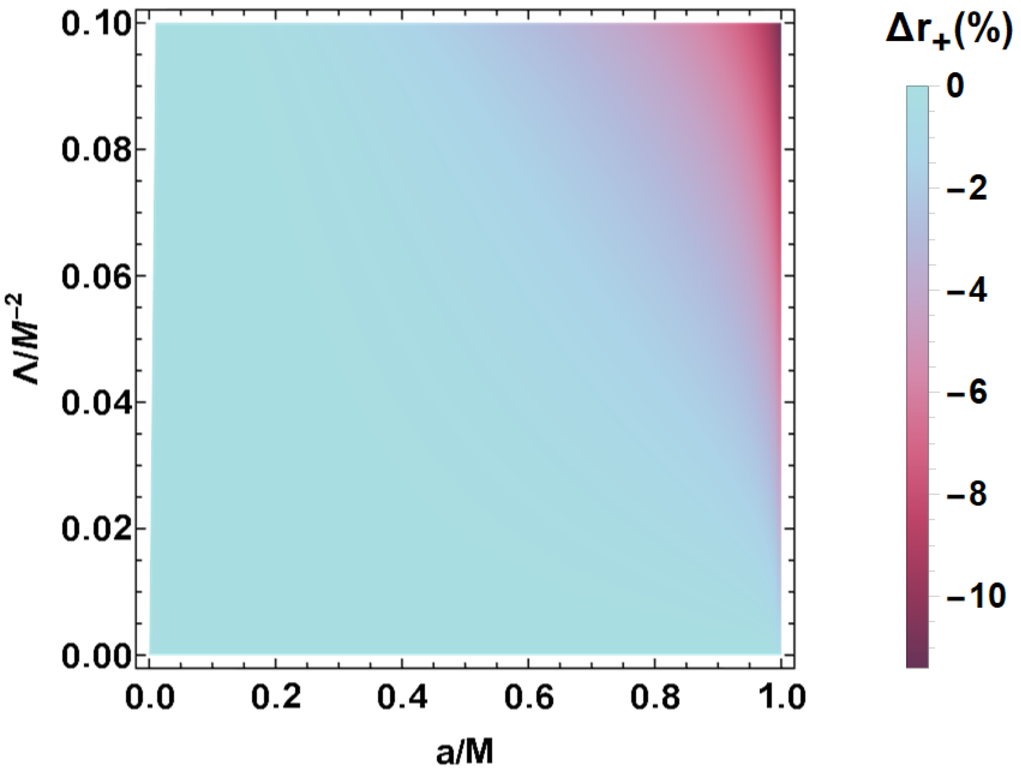}&
         \includegraphics[scale=0.85]{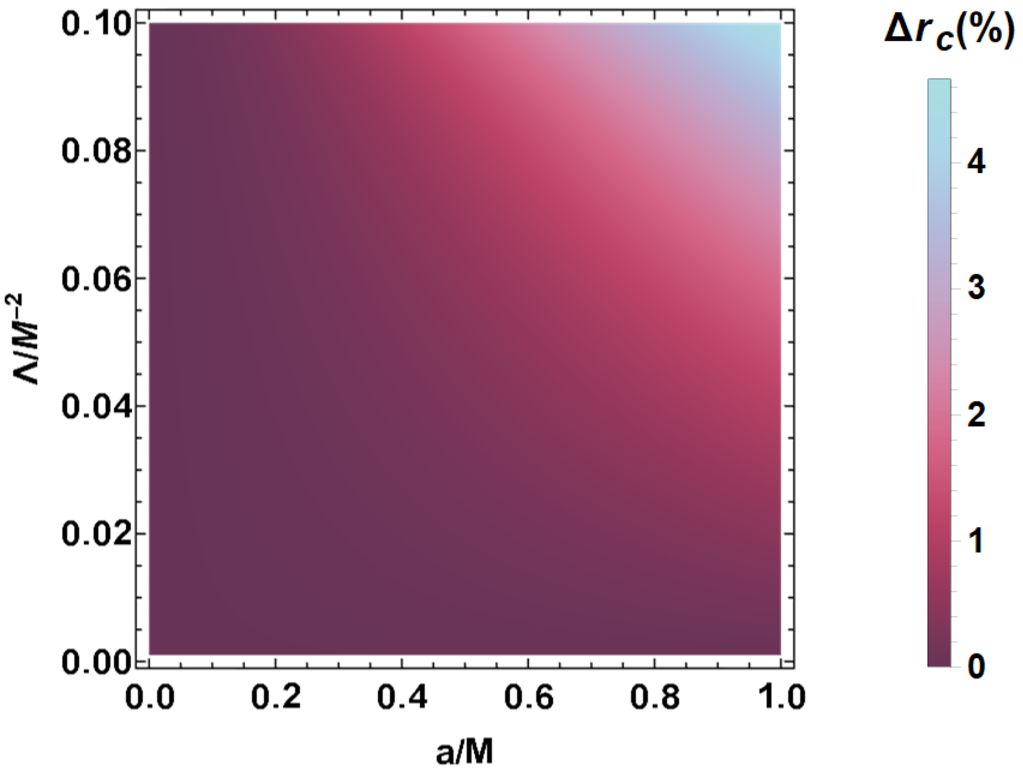}
    \end{tabular}
\end{adjustwidth}
\caption{Density plot of the percentage deviation of the event horizon ({\textbf{{left}}}) and cosmological horizon ({\textbf{right}}) of the new Kerr--de Sitter black holes (\ref{metric}) from that of the  Kerr--de Sitter black hole shown in parameter space.}\label{horizons_deviation}
\end{figure}

The metric (\ref{metric_ original}) is singular at the points where $\Sigma\neq0$ and $\Delta_r=0$ is a surface, namely event horizon. However, $\Delta_r=0$, for a given $M$, $a$ and $\Lambda$ admit three positive roots $r_{-}$, $r_+$ and $r_c$ such that $r_{-}<r_+<r_c$, respectively, correspond to Cauchy horizon ($r_{-}$),  event horizon ($r_+$) and the outermost cosmological horizon ($r_c$) \citep{Bhattacharya:2017scw}. When $\Lambda=0$, the Kerr horizons become
$
r_{\pm}^{k} = M\pm\sqrt{M^2-a^2}.
$
The area of the event horizon ($r_{+}$) is: $$A=\frac{4\pi(r_+
^2+a^2)}{\Xi}.$$

On the other hand, the entropy, temperature and angular velocity in terms of $r_+$ are, respectively \citep{Gibbons:1977mu}:
\vspace{6pt}

 $$S=\frac{A}{4},\; T=\frac{r_+(1-\frac{\Lambda a^2}{3}-\Lambda r_+- \frac{a^2}{r_+^2})}{4\pi (r_+^2+a^2)},\;\Omega=\frac{a(1-\frac{\Lambda r_+^2}{3})}{r_+^2+a^2}.$$

However, in the above, the chosen horizon can be Cauchy, event or cosmological, the Kerr--de Sitter black holes satisfy: $$dM=TdS_+ +\Omega_+ dJ$$ and this is true for each horizon. Furthermore, for $\Lambda<0$ and $\Lambda=0$, respectively, the above quantities are associated with the Kerr--anti-de Sitter and Kerr black holes. Similarly to the Kerr black hole, the Kerr--de Sitter black holes have killing vectors $\delta _t^{\mu }$  and $\delta _{\phi }^{\mu }$ corresponding to cyclic coordinates $t$ and $\phi$.
\subsection{A New Kerr--de Sitter Black Hole}
The Kerr--de Sitter solution of Carter \citep{Carter:1973rla} leaves the cosmological constant $\Lambda$ immaculate. However, the $\Lambda$ in the strong field regime, such as in the vicinity of a black hole, is no longer constant in quantum field theory \citep{Ovalle:2021jzf}. Recently, Ovalle et al. \citep{Ovalle:2021jzf} interpreted the cosmological constant $\Lambda$ as \textit{vacuum energy}  and implemented the so-called gravitational decoupling (GD) approach \citep{Ovalle:2017fgl,Ovalle:2018gic,Contreras:2021yxe} to  obtain a new Kerr--de Sitter black hole solution which is geometrically richer and has an impact of rotation in the form of a warped curvature \citep{Ovalle:2021jzf}.
The new Kerr--de Sitter black holes in the Boyer--Lindquist coordinates read \citep{Ovalle:2021jzf}:
\begin{align}\label{metric}
ds^{2}=-&\left[\frac{\Delta-a^{2}\sin^{2}\theta}{\Sigma}\right]dt^{2}+\frac{\Sigma}{\Delta}dr^{2}+\Sigma d\theta^{2}
-2a\sin^2{\theta}\left[1-\frac{\Delta-a^2\sin^2{\theta}}{\Sigma}\right]dtd\phi\nonumber\\
 +&\frac{\sin^2{\theta} }{\Sigma}\left[(r^2+a^2)^2-\Delta a^2\sin^2{\theta} \right]d\phi^{2},
\end{align}
where $\Delta=r^{2}+a^{2}-2Mr-{(\Lambda r^{4})}/{3}$ and $\Sigma=r^{2}+a^{2}\cos^{2}\theta$. The metric (\ref{metric}) describes the rotating black holes in asymptotically de Sitter or anti-de Sitter, respectively, when $\Lambda>0$ or $\Lambda<0$ and encompass the Kerr black hole in the absence of the cosmological constant ($\Lambda=0$) and Schwarzschild--de Sitter black hole ($a=0$) \citep{Ovalle:2017fgl}.
When written in Boyer--Lindquist coordinates, metric (\ref{metric}) is same as the Kerr black hole with $M$ replaced by $m(r)=M+\frac{\Lambda}{6}r^{3}$.
Thus, the new Kerr--de Sitter black hole is different from the Kerr--de Sitter black hole (\ref{metric_ original}), which is a $\Lambda$-vacuum solution.
In contrast, the new Kerr--de Sitter metric (\ref{metric}) is not a $\Lambda$-vacuum solution and has a curvature warped in space given by \citep{Ovalle:2021jzf}:
\begin{equation}
    R (r,\theta)=-4\Lambda\left[\frac{r^2}{r^2+a^2\cos^2{\theta}}\right],
\end{equation}
which is obeyed by the Carter's Kerr--de Sitter metric in the equatorial plane ($\theta=\pi/2$). Further, the deformation in $R$ becomes most prominent as $r\sim a$ and vanishes for $r>>a$ (cf.  Figure~\ref{Curvature}).
\begin{figure}
\begin{adjustwidth}{-\extralength}{0cm}
\centering
    \begin{tabular}{c c c }
	\includegraphics[scale=0.66]{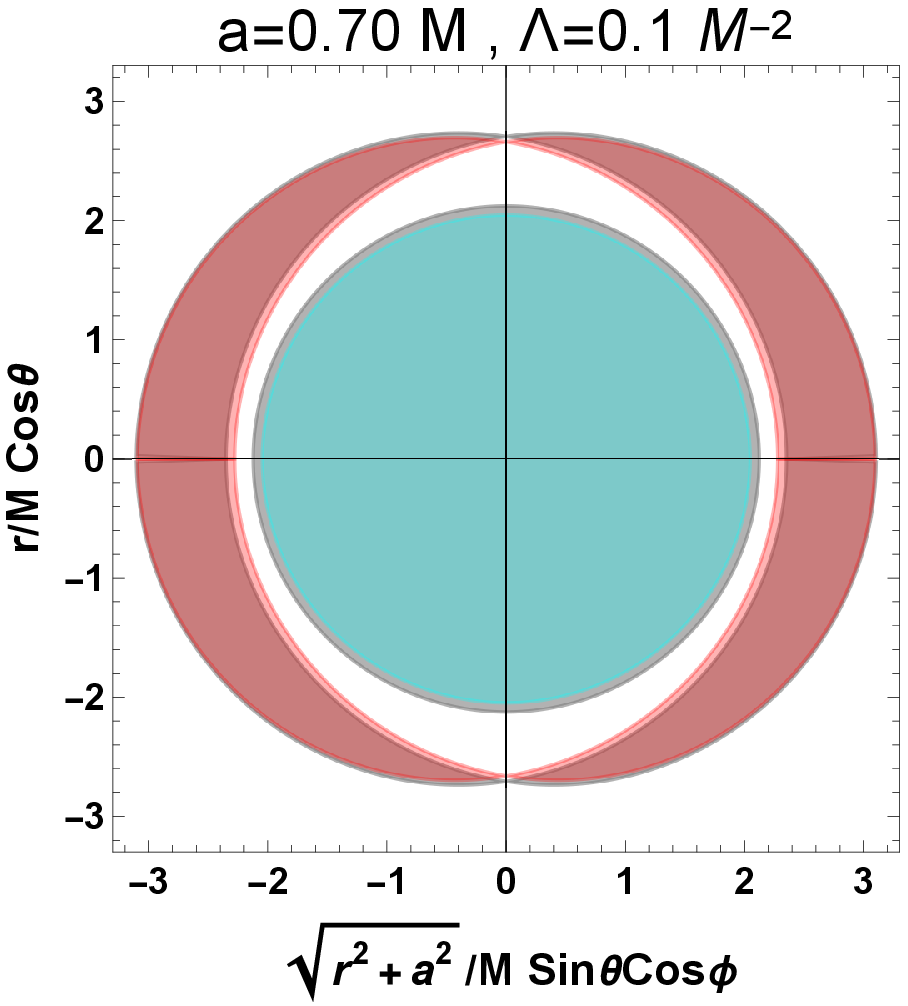}&
	\includegraphics[scale=0.66]{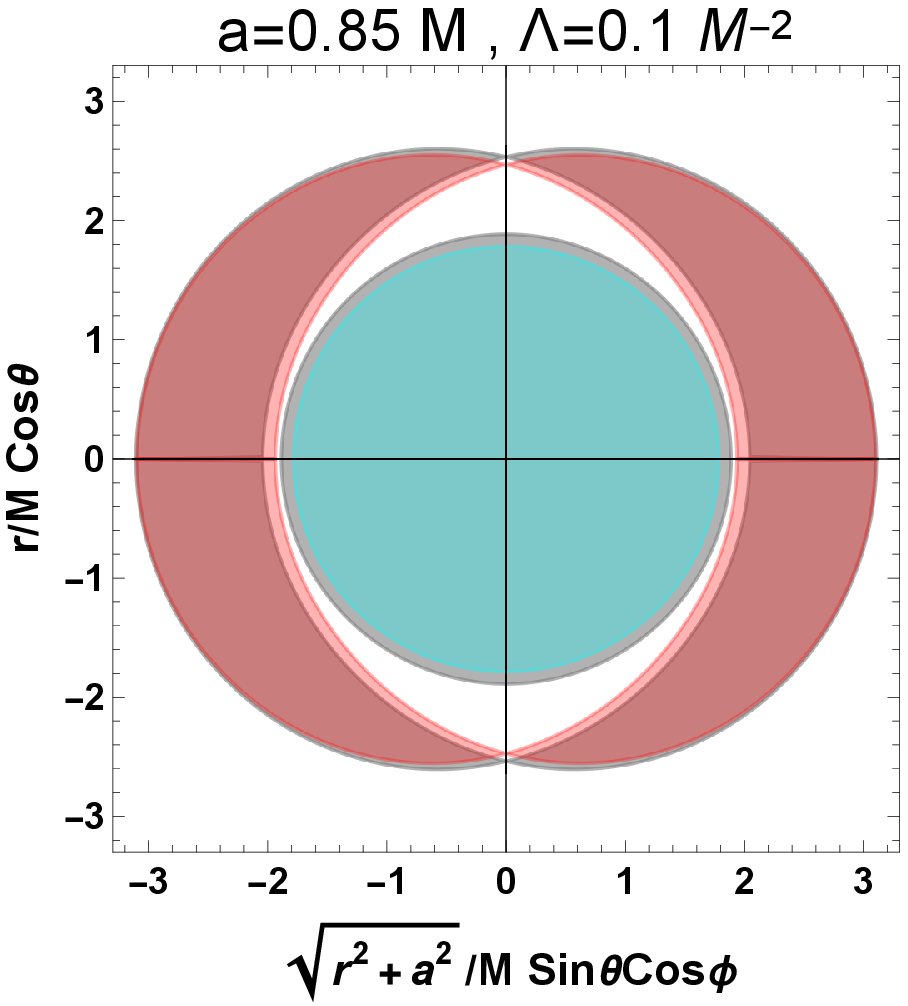}&
	\includegraphics[scale=0.66]{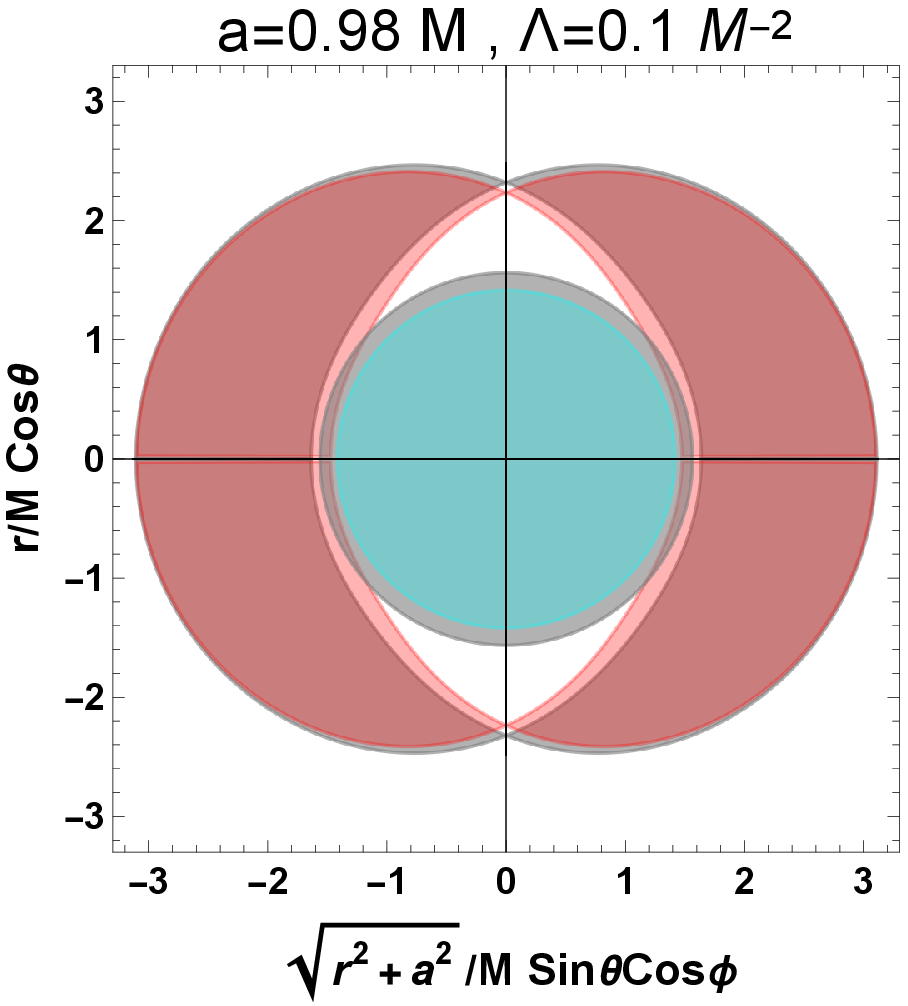}
\end{tabular}
\end{adjustwidth}
\caption{A comparison of the cross-section of the event horizon (blue shaded region) and photon region (pink shaded region) for the two Kerr--de Sitter black holes. The lighter regions correspond to the new Kerr--de Sitter black holes, whereas the darker regions correspond to the  Kerr--de Sitter black holes (\ref{metric_ original}).}\label{photonRegion_plot}	
\end{figure}
\paragraph{Energy Conditions}
If $T_{\mu\nu}$ is the energy momentum associated with matter for metric (\ref{metric}), we can define the components of the energy momentum tensor which in the orthonormal frame reads~\citep{Bambi:2013ufa,Neves:2014aba}:
\begin{equation}
T^{(a)(b)}=e_{\mu}^{(a)}e_{\nu}^{(b)} G^{\mu\nu},
\end{equation}
where $e_{\mu}^{(a)}$ is the basis for the usual orthonormal tetrad \citep{Bambi:2013ufa,Ghosh:2014pba,Ghosh:2015ovj,Neves:2014aba}.
We can write the components of the energy momentum tensor as \citep{Ovalle:2021jzf}
\vspace{6pt}

\begin{eqnarray}
\label{energyax}
{\rho}&=&-{p}_{r}=\frac{2\,r^2}{\Sigma ^2}\, {m}'=\Lambda\,\frac{r^4}{\Sigma^2}=\frac{\tilde{\Lambda}^2}{\Lambda}\ ,\\	\label{pressuresax}
{p}_{\theta}&=&{p}_{\phi}=-\frac{r }{\Sigma}\,{m}''+\frac{2\left(r^2-\Sigma\right)}{\Sigma ^2}\,{m}'=\rho-2\,\Lambda\,\frac{r^2}{\Sigma}\ .
\end{eqnarray}

The weak energy condition \citep{Bambi:2013ufa,Ghosh:2014pba,Ghosh:2015ovj,Ghosh:2021clx,Neves:2014aba} requires $\rho\geq0$ and $\rho+P_{i}\geq0$ ($i=r, \theta, \phi$). Obviously:
\begin{equation}
\rho+P_\theta=\rho+P_\phi=-\frac{\Lambda r^2(2\Sigma-r^2)}{\Sigma^2}.\label{rho_plus_P}
\end{equation}

Clearly, the weak energy conditions are satisfied for $\Lambda<0$.  However, in general,  as seen from Figure~\ref{rho_P_plot}, it may not hold. On the other hand, the dominant energy condition demands that:
\begin{equation}
\rho\geq0,\, \rho\geq|{P}_{i}|\quad\left(i=r,\theta,\phi\right).
\end{equation}

The dominant energy condition holds for $\Lambda>0$, but is found to be violated for \mbox{$\Lambda<0$ \citep{Ovalle:2021jzf}}. Furthermore, the strong energy condition given by
\begin{equation}
\rho+{P}_{r}+2\,{p}_{\theta}\geq 0,\, {\rho}+{p}_{r}\geq 0,\, {\rho}+{p}_{\theta}	\geq 0
\end{equation}
is satisfied for $\Lambda<0$ but is violated for $\Lambda>0$ \citep{Ovalle:2021jzf}. Our results show that the energy conditions may not be prevented, e.g., cf. Figure~\ref{rho_P_plot}. The weak energy condition is not satisfied for Kerr--de Sitter black holes, but the violation can be minimal, depending on the value of parameters, as shown in Figure~\ref{rho_P_plot}. Despite violations, such solutions are essential from a phenomenological point of view, and  they are also essential as astrophysical black holes are rotating \citep{Bambi:2013ufa,Ghosh:2014pba,Ghosh:2015ovj,Ghosh:2021clx,Neves:2014aba}.

Similar to the Kerr--de Sitter metric (\ref{metric_ original}), $\Sigma\neq0$ and $g^{rr}=\Delta=0$ give the following four roots \citep{Ovalle:2017fgl}:
\begin{eqnarray}
\label{roots}
&&r_{c}^{NK}=\frac{\sqrt{\kappa}}{2}+\frac{1}{2}\sqrt{\frac{6}{\Lambda}-\kappa-\frac{12M}{\Lambda\sqrt{\kappa}}}\ ,\;\;
r_{+}^{NK}=\frac{\sqrt{\kappa}}{2}-\frac{1}{2}\sqrt{\frac{6}{\Lambda}-\kappa-\frac{12M}{\Lambda\sqrt{\kappa}}}\ ,\nonumber\\
&&r_{-}^{NK}=-\frac{\sqrt{\kappa}}{2}+\frac{1}{2}\sqrt{\frac{6}{\Lambda}+\kappa+\frac{12M}{\Lambda\sqrt{\kappa}}}\ ,\;\;
r_{--}^{NK}=-\frac{\sqrt{\kappa}}{2}-\frac{1}{2}\sqrt{\frac{6}{\Lambda}-\kappa+\frac{12M}{\Lambda\sqrt{\kappa}}}\ ,
\end{eqnarray}
where $\kappa={(2\omega+\delta+\omega^2)}/{(\Lambda\omega)}$, $\delta=1-4a^2\Lambda$, $\omega=(Q-P)^{\frac{1}{3}}$, $P=1+12a^2\Lambda-18M^2\Lambda$, $Q=2[a^2\Lambda(3+4a^2\Lambda)^2+9M^2\Lambda(9M^2\Lambda-12a^2\Lambda-1)]^{\frac{1}{2}}$.
The above roots correspond, respectively, to the cosmological horizon, the event horizon, the Cauchy horizon and the inner cosmological horizon (inside the singularity) such that
$r_{--}^{NK}<r_{-}^{NK}<r_{+}^{NK}<r_{c}^{NK}$.

The  Figure~\ref{horizon_plot} shows the behaviour of the event horizon ($r_{+}^{NK}$) and the cosmological horizon ($r_{c}^{NK}$) of the new Kerr--de Sitter black holes in the ($a/M-\Lambda/M^{-2}$) space. The horizon structure of the new Kerr--de Sitter black holes is found to be substantially different from the Kerr--de Sitter black hole which is evinced by the relative difference of the horizons of the two black holes defined as:
\begin{align}
\Delta r_{+}=\frac{r_{+}^{NK}-r_{+}}{r_{+}}\;\;\;\text{and}\;\;\;\Delta r_{c}=\frac{r_{c}^{NK}-r_{c}}{r_{c}}.
\end{align}

Interestingly, $r_{+}^{NK}\leq r_{+}$ and $r_{c}^{NK}\geq r_{c}$, for the entire ($a/M$-$\Lambda /M^{-2}$) space and while there is a maximum of 11.408\% negative deviation of the event horizon of the new Kerr--de Sitter black holes, the cosmological horizon deviates by a maximum of 4.656\% (cf.  Figure~\ref{horizons_deviation}); thus, a larger screening effect due to the $\Lambda$ is observed in the new Kerr--de Sitter black holes.

For $\Lambda>0$, the region between the event horizon ($r_{+}^{NK}$) and cosmological horizon ($r_{c}^{NK}$) is called the \textit{domain of outer communication} where $g_{rr}>0$ leads to $\partial_r$ being space-like and thus communication between two observers is possible in this region \citep{Kumar:2017tdw}. In the domain of outer communication, the killing vector field $\partial_t$ is time-like and the \mbox{spacetime (\ref{metric})} being static allows the presence of static observers in this region. No static observer can be considered in the regions $r<r_{+}^{NK}$ or $r>r_{c}^{NK}$ wherein the causal character of vector field $\partial_r$ changes from space-like to time-like, and thus, such an observer would be invisible to someone in the domain of outer communication \citep{Kumar:2017tdw}.
The cosmic expansion will drive away an observer outside the $r_{c}^{NK}$, whereas one inside the $r_{+}^{NK}$ would fall into the singularity \citep{Perlick:2018iye}. Herein, we shall consider a static observer placed at a radial coordinate  $r_{+}^{NK}<r_o<r_{c}^{NK}$, particularly, at Boyer--Lindquist coordinates ($t_o, r_o,\theta_o=\pi/2,\phi_o=0$) \citep{Kumar:2017tdw}. For an observer at $r_{+}^{NK}<r_o<r_{c}^{NK}$, $r_{+}^{NK}$ is the future inner horizon while $r_{c}^{NK}$ is the future outer horizon, and crossing any of these two horizons will break the causal connection with region $r_{+}^{NK}<r<r_{c}^{NK}$ \citep{Kumar:2017tdw}. 
  \begin{figure}
\begin{adjustwidth}{-\extralength}{0cm}
\centering
    \begin{tabular}{c c}
	\includegraphics[scale=0.75]{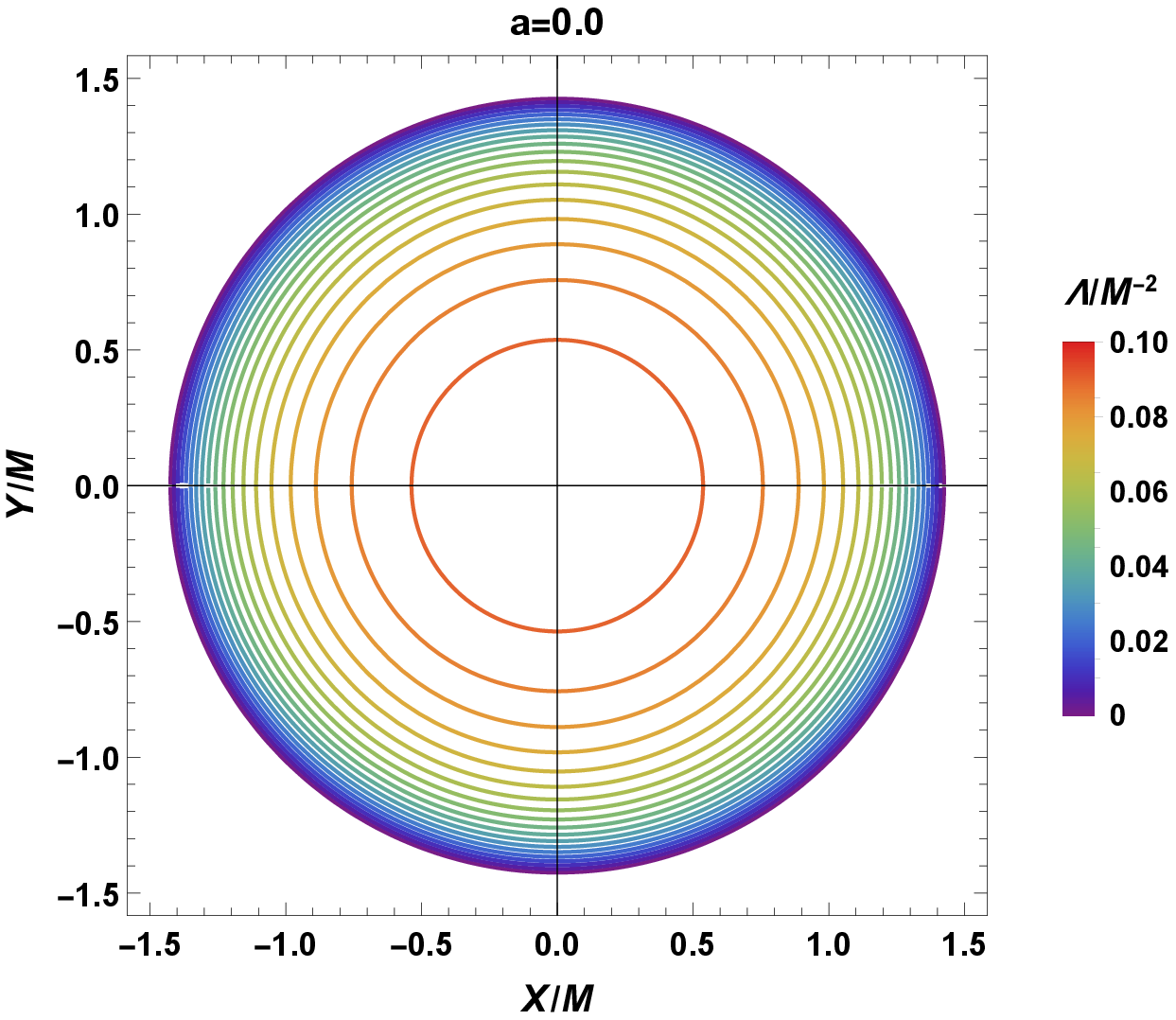}&
	\includegraphics[scale=0.75]{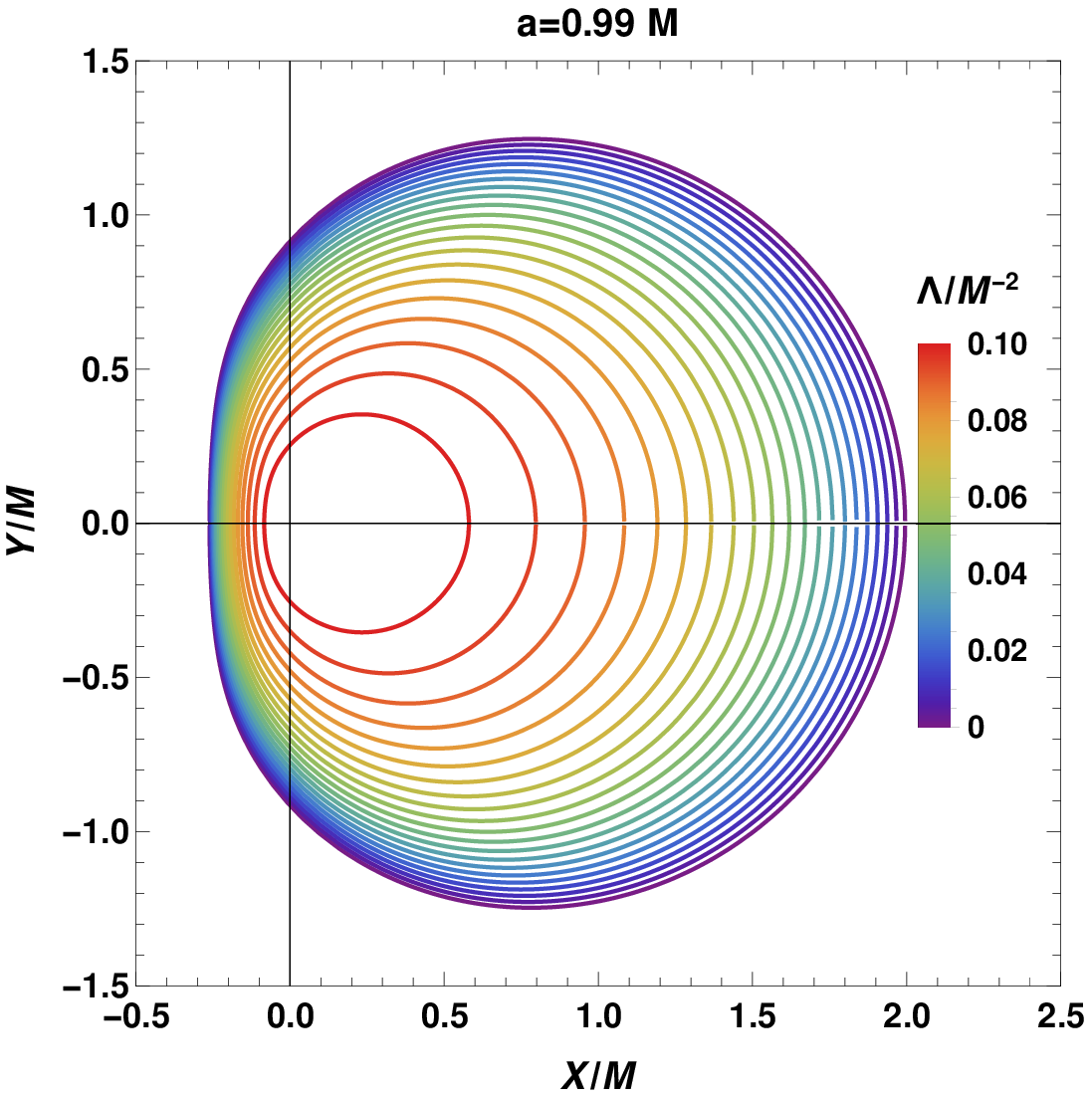}\\
\end{tabular}
\end{adjustwidth}
\caption{Shadow profile of Schwarzschild--de Sitter black holes ({\textbf{left}}) and new Kerr--de Sitter black holes ({\textbf{right}}) with a varying  $\Lambda$ compared with Kerr black hole (outermost solid curve corresponding to $\Lambda \to 0$).}\label{shadow}	
\end{figure}
\section{Photon Region and Shadow in Asymptotic de Sitter Spacetime}\label{Sec3}

In order to discuss the photon region \citep{Teo:2020sey,Johnson:2019ljv} and black hole shadow \citep{Kumar:2020hgm,Kumar:2019pjp,Kumar:2019ohr,Ghosh:2019eoo} in the new Kerr--de Sitter spacetime, it is pertinent to study the motion of a test particle. We consider the black hole in a luminous background and the photons from a source, due to gravitational lensing \citep{Kumar:2020sag,Islam:2020xmy,Kumar:2020owy}, get deflected before reaching an observer. The photons, depending on the energy, may get scattered, captured or move in unstable orbits \citep{Chandrasekhar:1985kt,Kumar:2020yem}. Tracing back these photon trajectories we get the black hole shadow \citep{Grenzebach:2014fha,Kumar:2020owy,Kumar:2020hgm,Kumar:2019pjp,Kumar:2019ohr,Kumar:2017tdw,Afrin:2021imp}. Hence, we need to discuss the null geodesics for the new Kerr--de Sitter metric (\ref{metric_ original}).
 Owing to the axial and time translational symmetries alongside a hidden symmetry \citep{Carter:1968rr}, the  spacetime (\ref{metric}) gives four constants of motion along each null geodesic: the mass $g_{\mu\nu}p^{\mu}p^{\nu}=0$, total energy $E=-p . \partial_{t}$, angular momentum $\mathcal{L}_z=p . \partial_{\phi}$ and Carter constant $\mathcal{Q}$. We follow the Carter's separable method to solve the Hamilton-Jacobi equation to obtain the following equations of motion around the new Kerr--de Sitter metric in the first order differential form\citep{Carter:1968rr,Chandrasekhar:1985kt}:
\begin{align}
\Sigma \dot{t}=&\frac{r^2+a^2}{\Delta}(\mathcal{E}(r^2+a^2)-a\mathcal{L}_z)-a(a\mathcal{E}\sin^2{\theta}-\mathcal{L}_z),\\
\Sigma \dot{\phi}=&\frac{a}{\Delta}(\mathcal{E}(r^2+a^2)-a\mathcal{L}_z)-\left(a\mathcal{E}-\frac{\mathcal{L}_z}{\sin^2{\theta}}\right),\\
\Sigma^2 \dot{r}^2=&\Big((r^2+a^2)E-a \mathcal{L}_z \Big)^2-\Delta \Big({\mathcal{K}}+(a E- \mathcal{L})^2\Big)\equiv\mathcal{R}(r)\ ,\label{req} \\
\Sigma^2 \dot{\theta}^2=&\mathcal{K}-\left(\frac{{L_z}^2}{\sin^2\theta}-a^2 E^2 \right)\cos^2\theta\equiv\Theta(\theta)\ ,\label{theq}
\end{align}
where the separability constant $\mathcal{K}$ is related to the Carter constant $\mathcal{Q}$ through $\mathcal{Q}=\mathcal{K}+(a\mathcal{E}-\mathcal{L}_z)^2$ \citep{Chandrasekhar:1985kt}. Further, dimensionless impact parameters $\xi=\mathcal{L}_z/\mathcal{E}\text{,}\;\eta=\mathcal{K}/\mathcal{E}^2$ \citep{Chandrasekhar:1985kt} reduce the degrees of freedom of the photon geodesics from three to two.
\begin{figure}
\begin{adjustwidth}{-\extralength}{0cm}
\centering
    \begin{tabular}{c c}
    \includegraphics[scale=0.8]{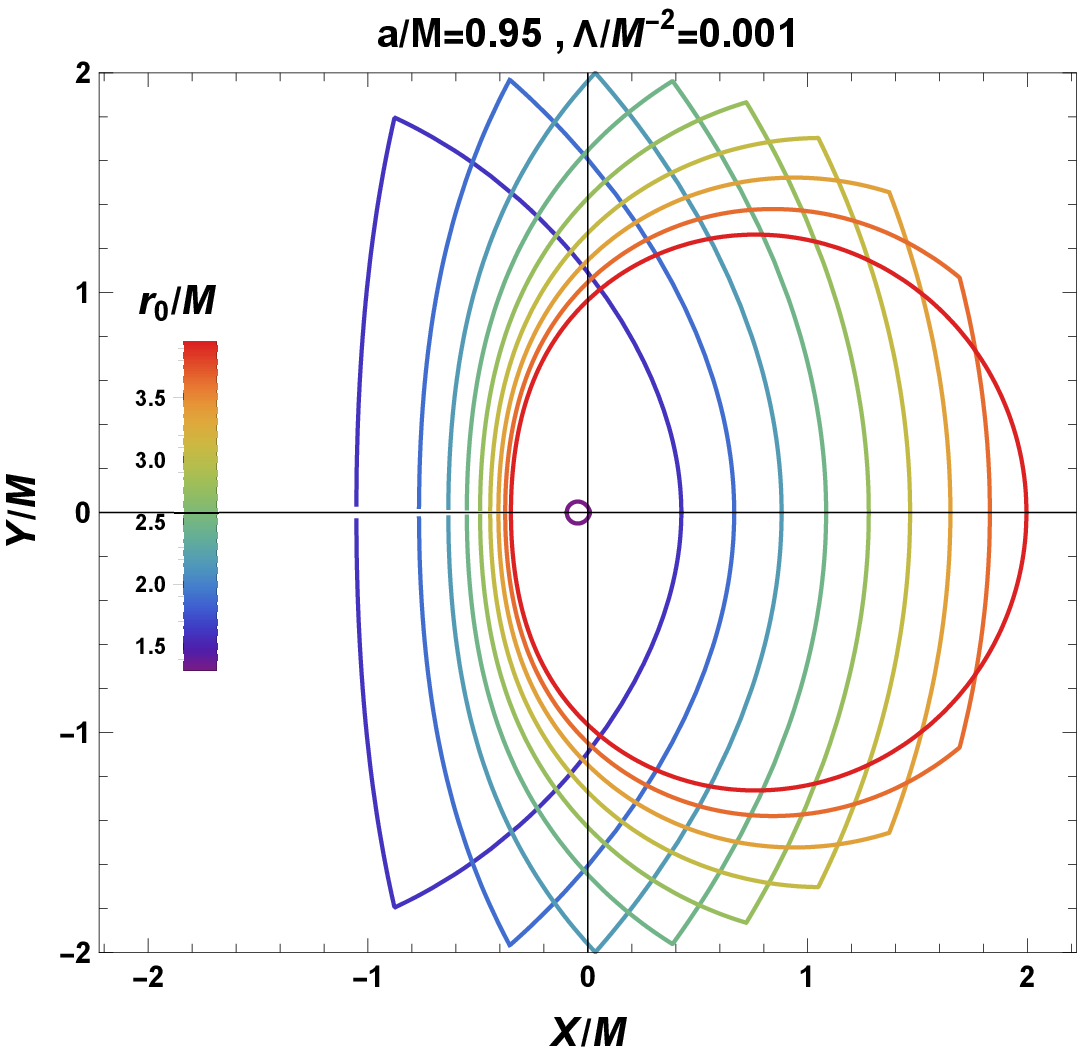}&
	\includegraphics[scale=0.8]{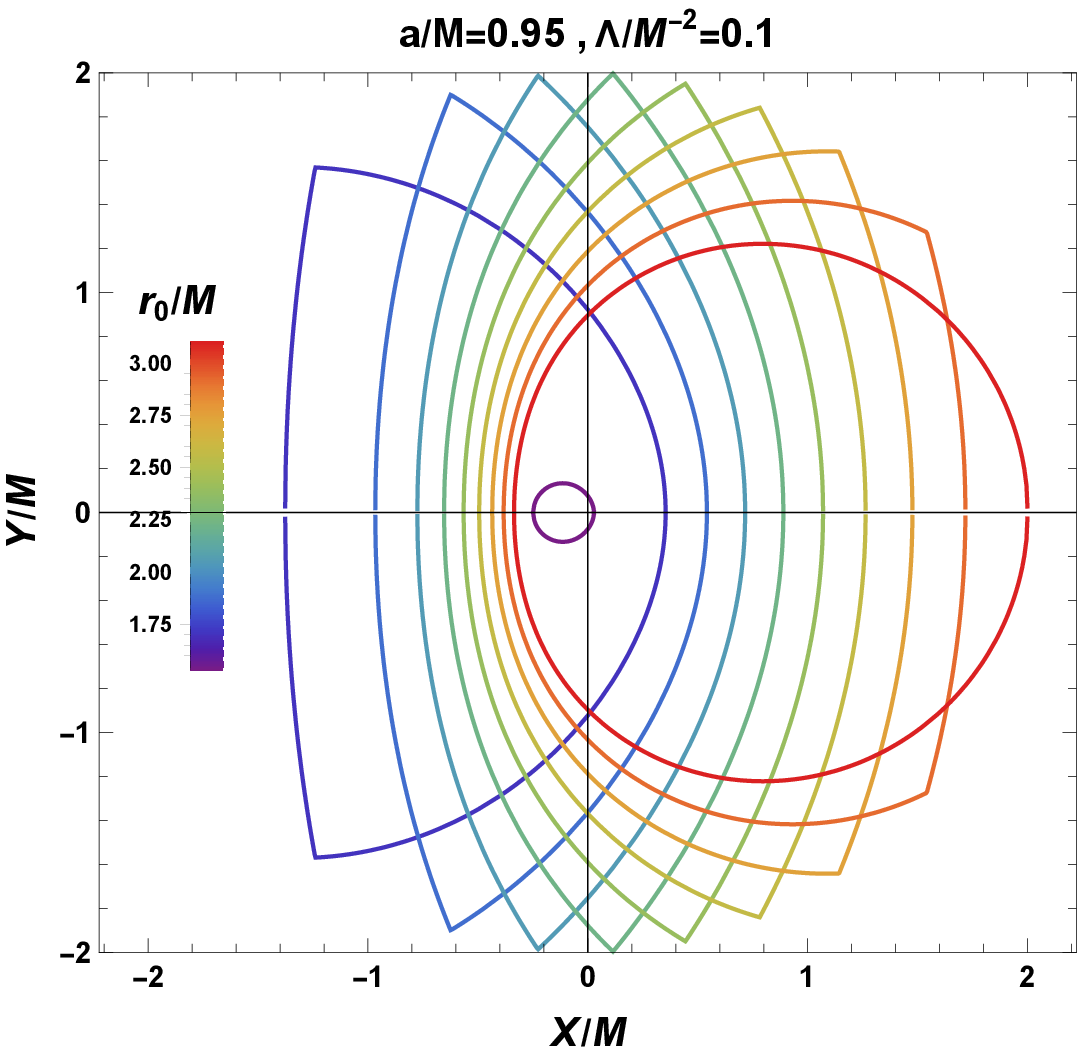}\\
	\includegraphics[scale=0.83]{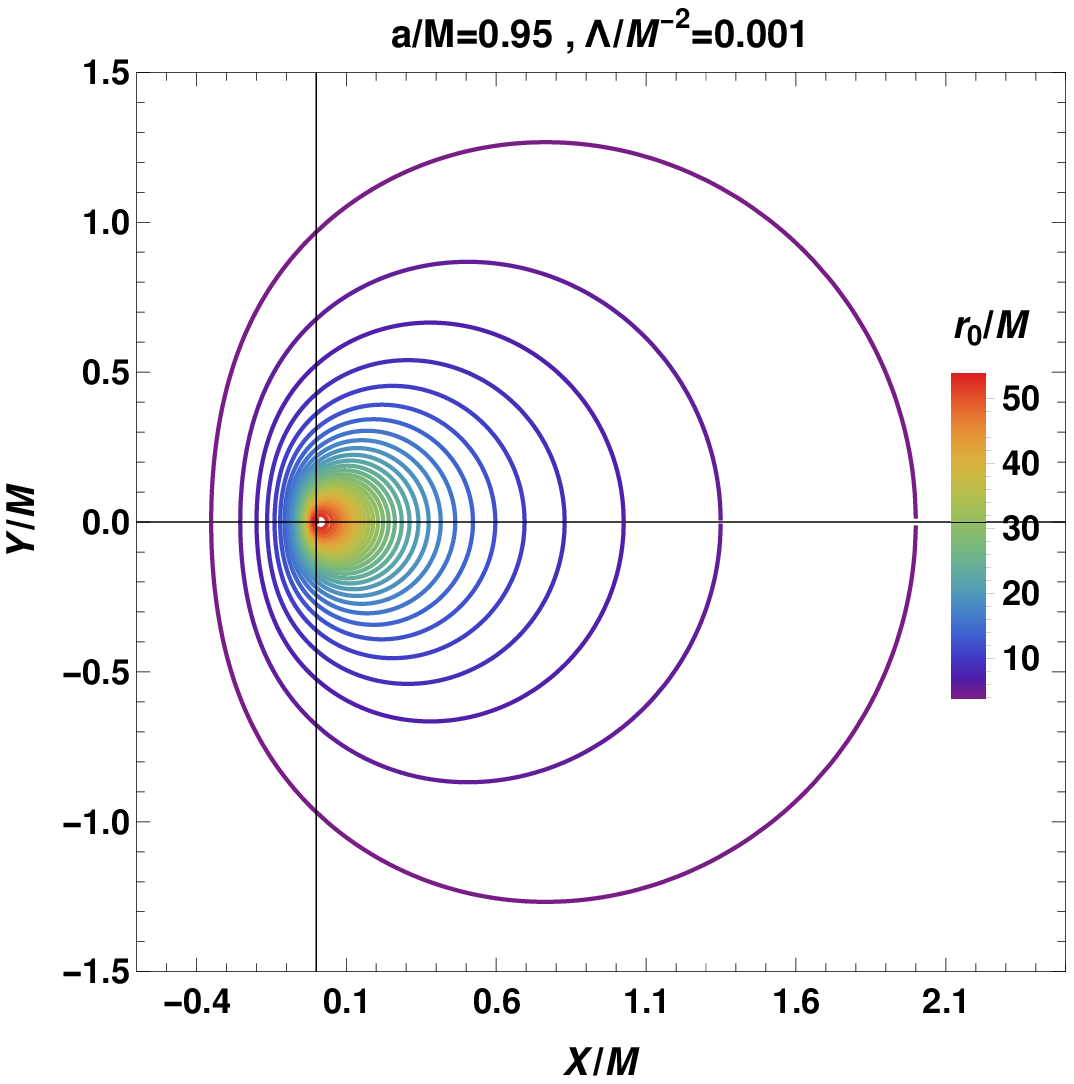}&
	\includegraphics[scale=0.83]{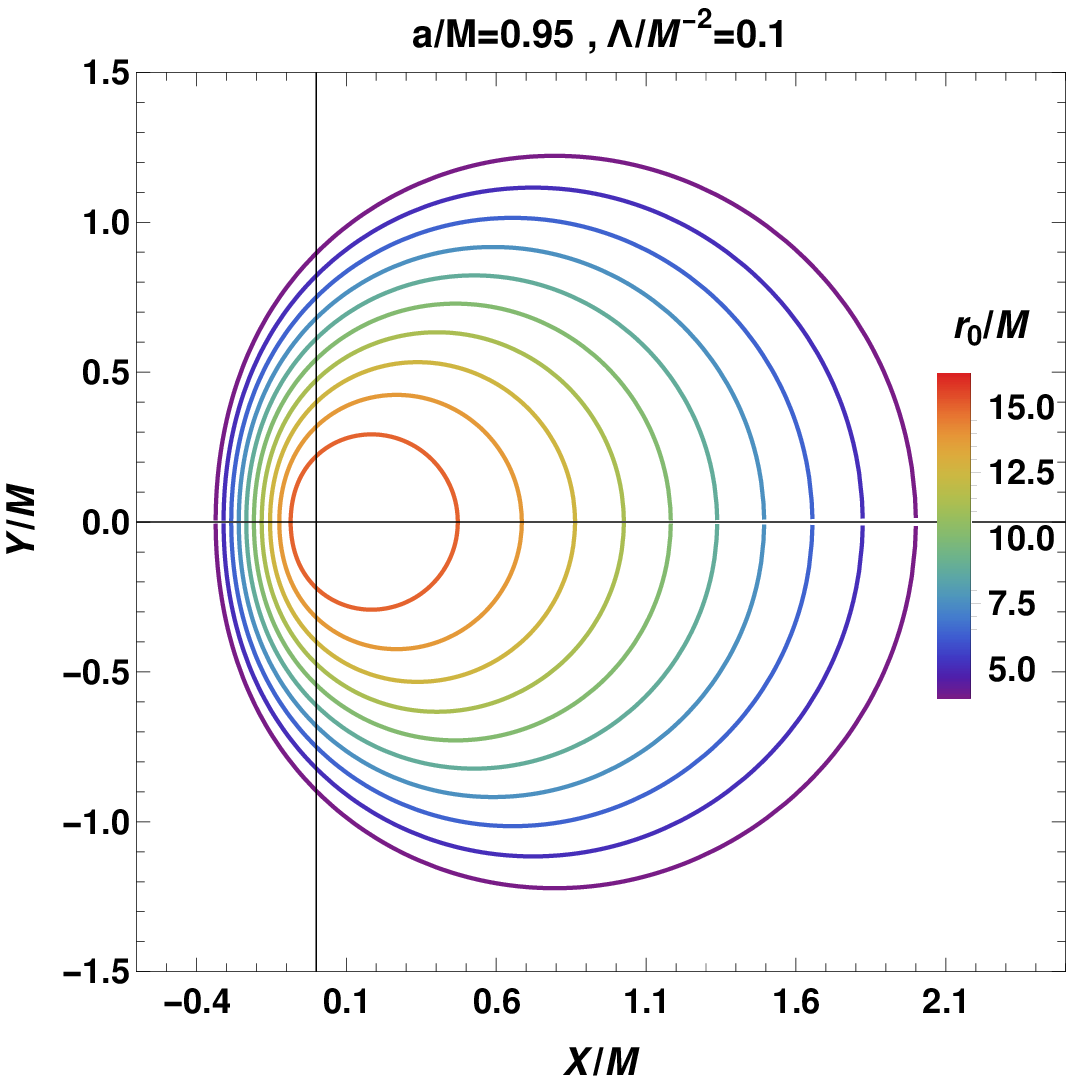}
\end{tabular}
\end{adjustwidth}
\caption{Shadow profile of  new Kerr--de Sitter black holes with varying observer distance\linebreak $r_{+}^{NK}<r_o<r_{p}^{+}$ ({\textbf{upper panels}}) and $r_{p}^{+}<r_o<r_{c}^{NK}$ ({\textbf{lower panels}}).}\label{shadow1}	
\end{figure}

It turns out that the unstable photon orbits outline the black hole shadow whose radii $r_p$ satisfy $\mathcal{R}(r_p)=\mathcal{R}'(r_p)=0$ \citep{Chandrasekhar:1985kt,Teo:2020sey,Kumar:2020yem}. The shadow boundary is the apparent shape of the unstable photon orbits due to strong gravitation lensing near the black hole.
We note that the Equation~(\ref{req}) can be recast as \citep{Kumar:2020yem}:
\begin{equation}
    \dot{r}^2+V_{eff}(r)=0,
\end{equation}
where $V_{eff}$ is the effective radial potential for the photons and is given by
\begin{equation}
    V_{eff}(r)=\frac{1}{\Sigma^2}\left[\Delta ({\mathcal{K}}+(a E- \mathcal{L})^2)-((r^2+a^2)E-a \mathcal{L}_z)^2\right].
\end{equation}
Thus, the unstable photon orbits have radial turning point when \citep{Kumar:2020yem}:
\begin{equation}
    V_{eff}(r)=\mathcal{R}=0\;\text{and}\;V_{eff}'(r)=\mathcal{R}'=0. \label{stability_eq}
\end{equation}
Solving Equation~(\ref{stability_eq}) for the impact parameters yield:
\begin{eqnarray}
  \xi_{crit}&=&\frac{a^2 \left(3 M+2 \Lambda  r_{p}^3+3 r_{p}\right)+3 r_{p}^2 (r_{p}-3 M)}{a\left(3 M+2 \Lambda  r_{p}^3-3 r_{p}\right)},\nonumber\\
\eta_{crit}&=&-\frac{3 r_{p}^3 \left[a^2 \left(4 \Lambda  r_{p}^3-12 M\right)+3 r_{p} (r_{p}-3 M)^2\right]}{a^2 \left(3 M+2 \Lambda  r_{p}^3-3 r\right)^2}\label{CriImpPara}.  
\end{eqnarray}
\begin{figure}[t]
\begin{adjustwidth}{-\extralength}{0cm}
\centering
    \begin{tabular}{c c}
        \includegraphics[scale=0.765]{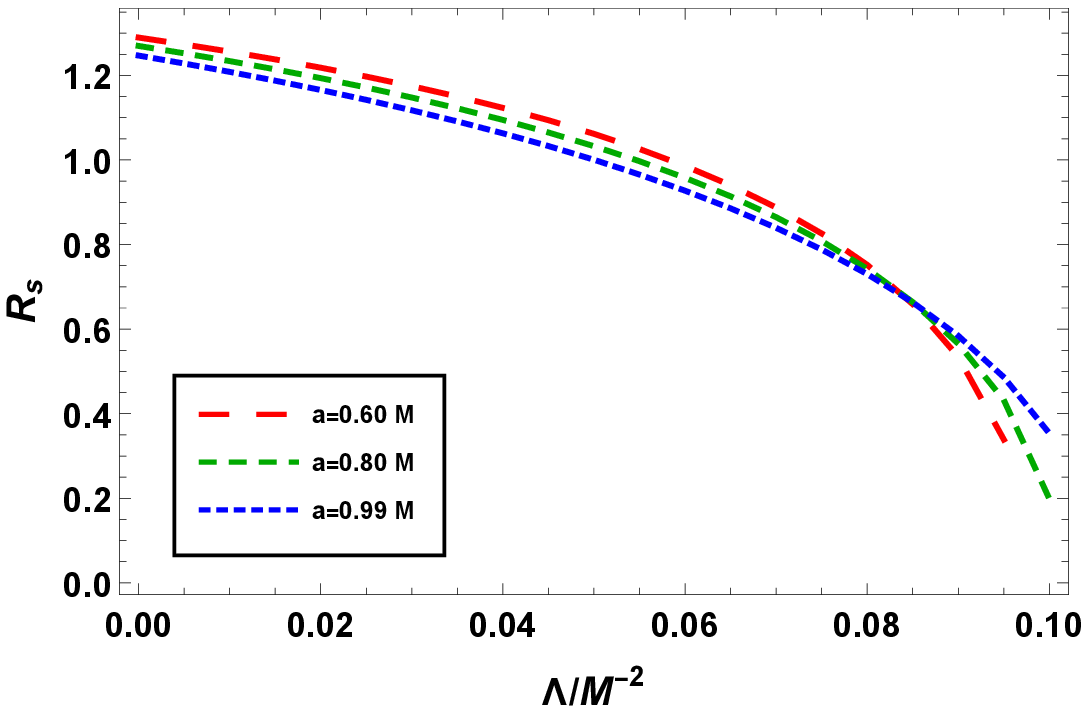}&
        \includegraphics[scale=0.75]{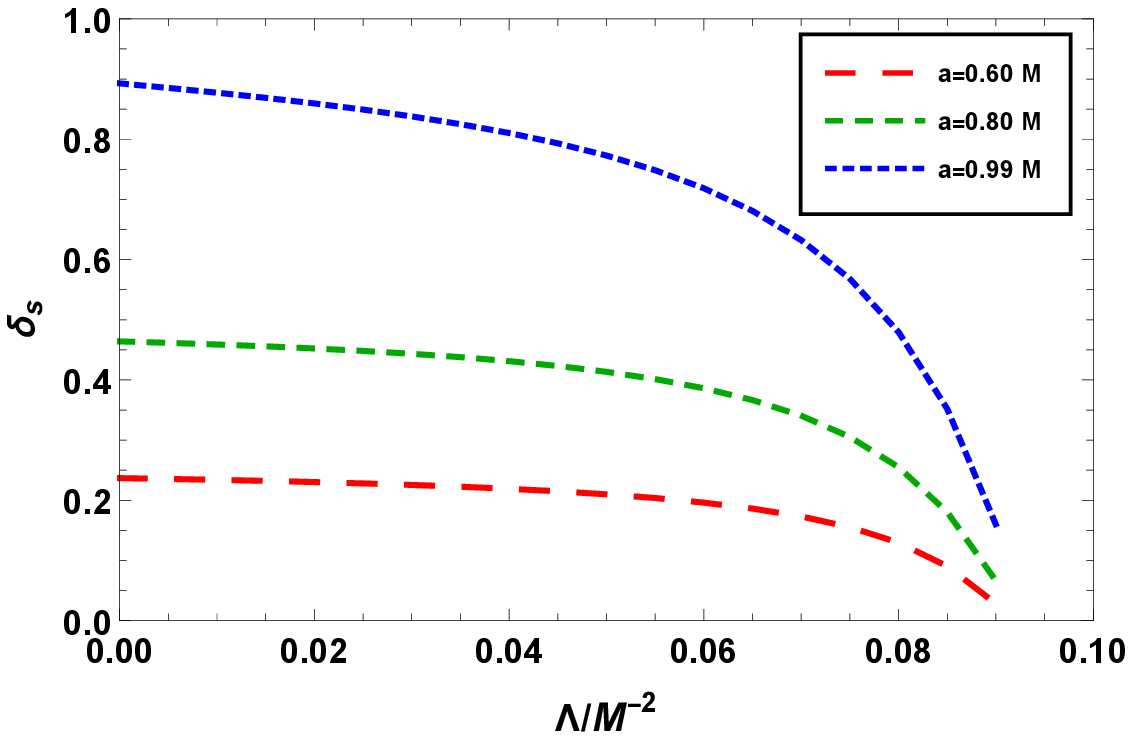}\\
        \includegraphics[scale=0.75]{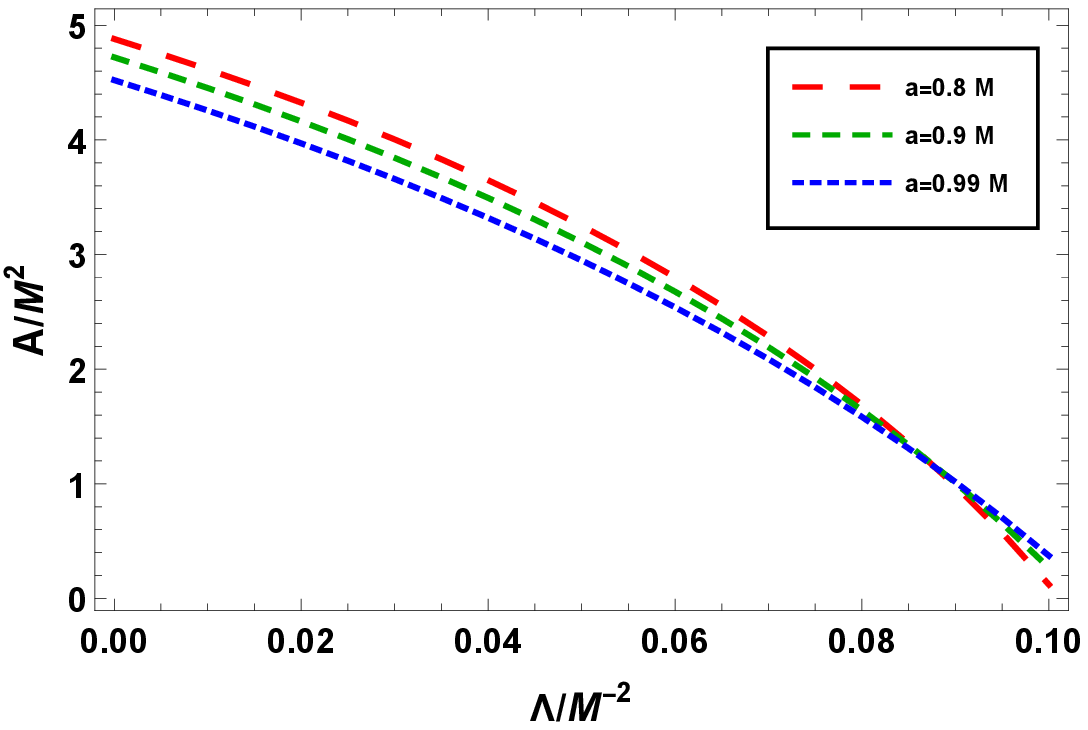}&
        \includegraphics[scale=0.782]{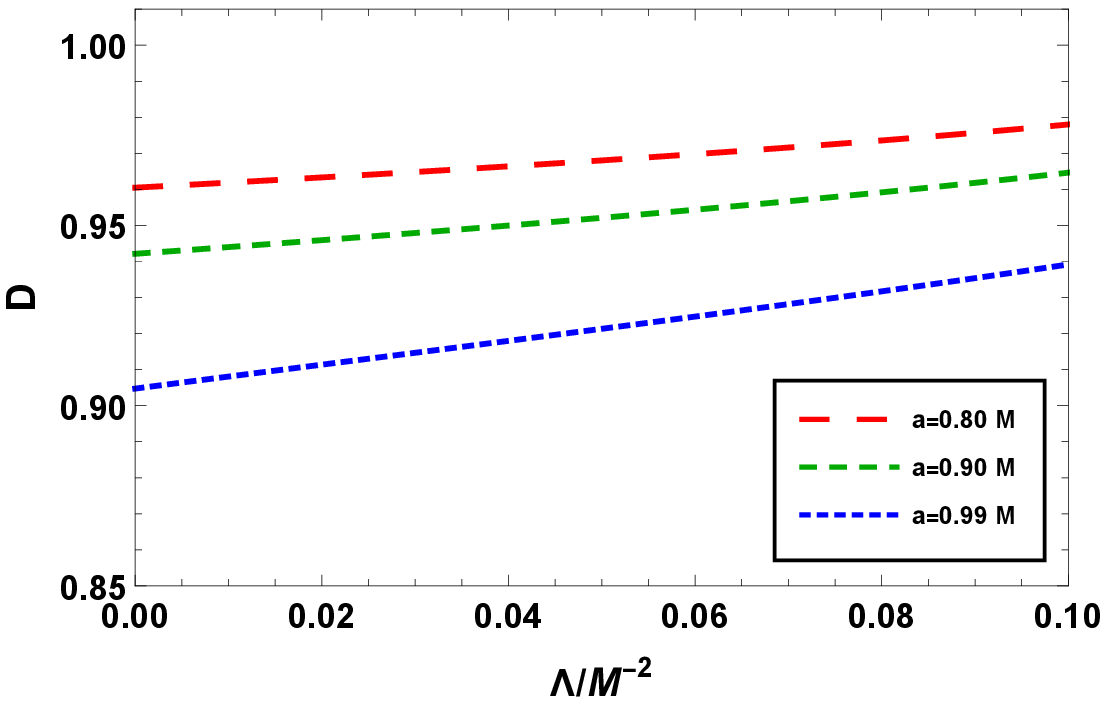}
    \end{tabular}	
\end{adjustwidth}
\caption{Variation of the shadow observables with $\Lambda$ for different spins $a$ for a static observer at $(r_o, \theta_o)=(4M, 90^{\circ})$.} \label{Observables_plot}
\end{figure}
\begin{figure}[t]
\begin{tabular}{c c}
	\hspace{-1.05cm}\includegraphics[scale=0.75]{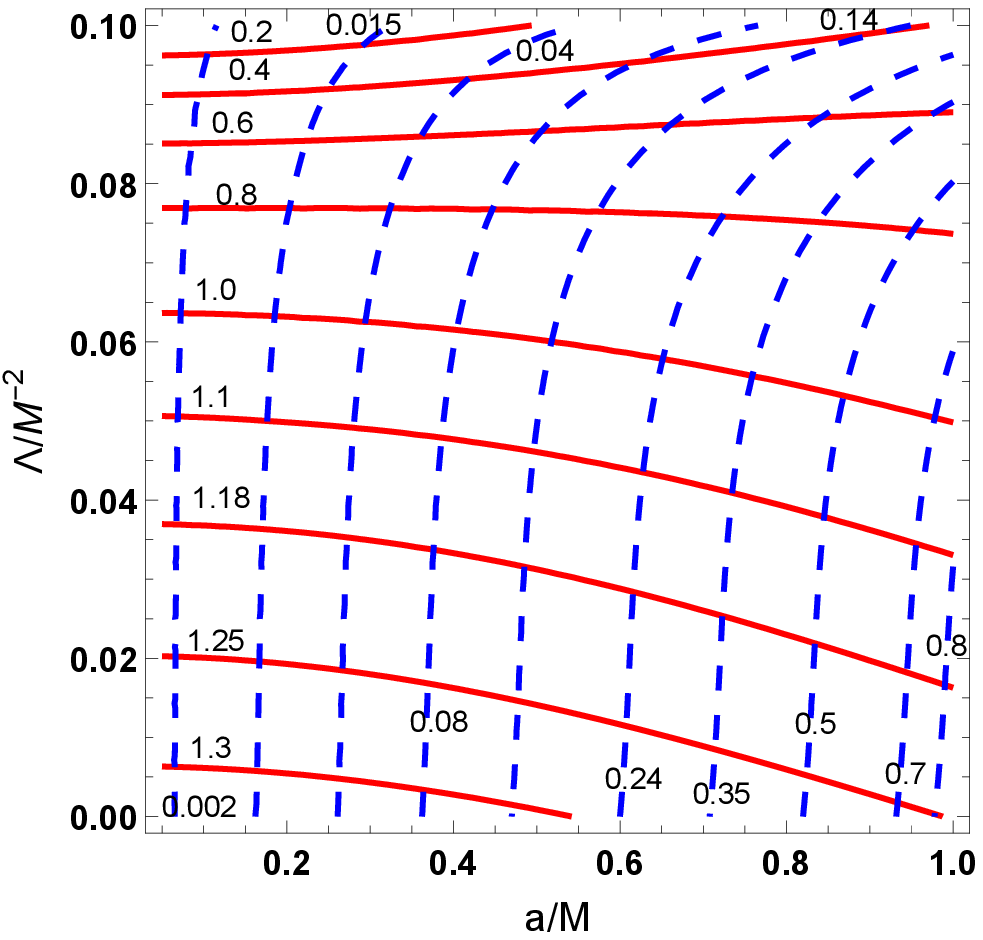}&
    \hspace{-0.3cm}\includegraphics[scale=0.79]{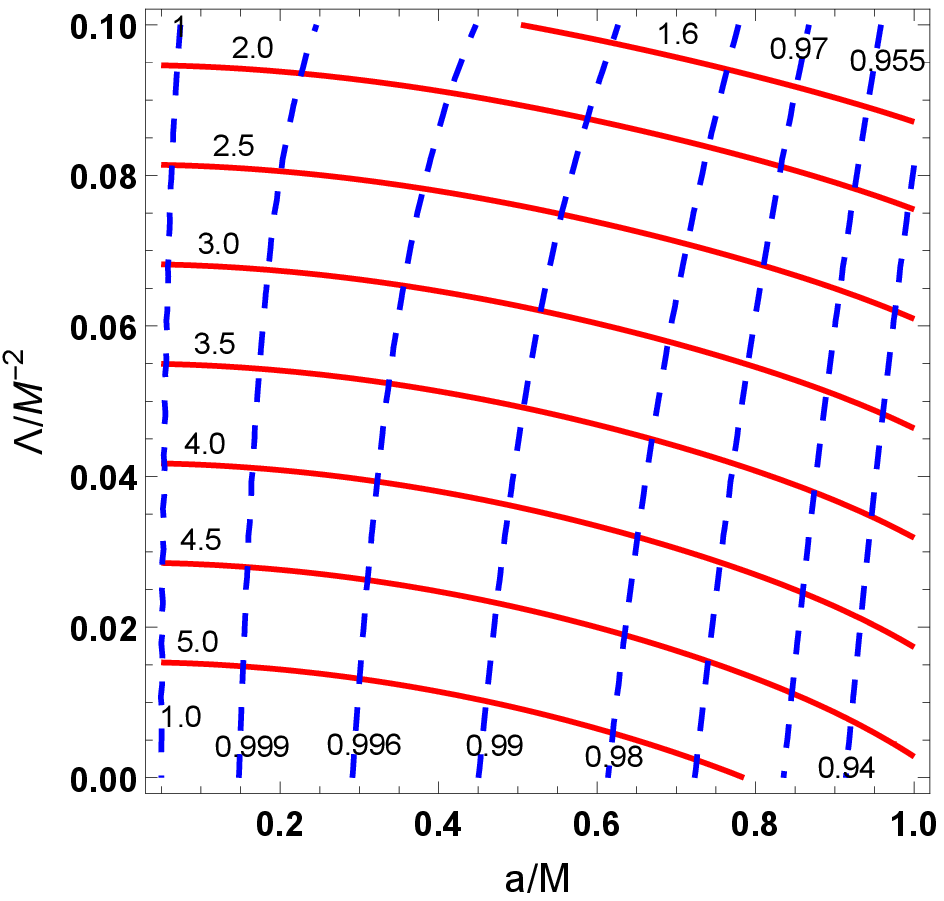}
	\end{tabular}
\caption{Contours for new Kerr--de Sitter black holes shadow observables,  $R_s$ (red solid lines) and $\delta_s$ (blue dashed lines) ({\textbf{left}}),  $A/M^2$ (red solid lines) and $D$ (blue dashed lines) ({\textbf{right}}) for a static observer at $(r_o, \theta_o)=(4M, 90^{\circ})$. The contour intersection points uniquely determine the black hole parameters. }\label{parameterEstimation_plot}
\end{figure}
\begin{table}[t]
\caption{Estimated parameters of new Kerr--de Sitter black holes.}
\centering
        \begin{tabular}{l  l  l l}
        \;\;\;\;$R_s$ & \;\;\;\;\;\;$\delta_s$ & \;\;\;\;$a/M$ & \;\;\;\;$\Lambda / M^{-2}$ \\ [0.8ex] 
 \hline\hline
 \;\; 1.30\;\; &\;\; 0.002\;\; &\;\; 0.06598\;\; &\;\; 0.00735\;\; \\   
 \;\; 1.25\;\; &\;\; 0.015\;\; &\;\; 0.16740\;\; &\;\; 0.02065\;\; \\    
 \;\; 1.10\;\; &\;\; 0.040\;\; &\;\; 0.27950\;\; &\;\; 0.05025\;\; \\   
 \;\; 0.80\;\; &\;\; 0.140\;\; &\;\; 0.57850\;\; &\;\; 0.07741\;\; \\   
 \;\; 0.60\;\; &\;\; 0.240\;\; &\;\; 0.85430\;\; &\;\; 0.08940\;\; \\   
 \hline
       \end{tabular}

            \centering

        \begin{tabular}{l  l  l l}
        \;\;$A/M^2$ & \;\;\;\;\;\;$D$ & \;\;\;\;$a/M$ & \;\;\;\;$\Lambda / M^{-2}$ \\ [0.8ex] 
 \hline\hline
 \;\; 5.0\;\; &\;\; 1.000\;\; &\;\; 0.05272\;\; &\;\; 0.01538\;\; \\   
 \;\; 4.0\;\; &\;\; 0.999\;\; &\;\; 0.16600\;\; &\;\; 0.04127\;\; \\   
 \;\; 3.5\;\; &\;\; 0.996\;\; &\;\; 0.34040\;\; &\;\; 0.05225\;\; \\    
 \;\; 2.0\;\; &\;\; 0.980\;\; &\;\; 0.74150\;\; &\;\; 0.08380\;\; \\   
 \;\; 1.6\;\; &\;\; 0.955\;\; &\;\; 0.94420\;\; &\;\; 0.08882\;\; \\   
 \hline
        \end{tabular}

     \label{Table_parameter}
\end{table}
We note that, the critical parameters for the  Kerr--de Sitter spacetime are slightly different from Equation~(\ref{CriImpPara}) and are given by
\begin{align}
    \xi_{crit}^{k}=&\frac{a^2 \left(3 M+\Lambda  r_{p}^3+3 r_{p}\right)+a^4 \Lambda  r_{p}+3 r_{p}^2 (r_{p}-3 M)}{a \left(r_{p} \left(a^2 \Lambda +2 \Lambda  r_{p}^2-3\right)+3 M\right)},\nonumber\\
\eta_{crit}^{k}=&-\frac{r^3 [6 a^2 \left(3 M \left(\Lambda  r_{p}^2-2\right)+\Lambda  r_{p}^3\right)+a^4 \Lambda ^2 r_{p}^3+9r_{p}(r_{p}-3 M)^2 ]}{a^2 \left[r_{p} \left(a^2 \Lambda +2 \Lambda  r_{p}^2-3\right)+3 M\right]^2}.\label{CriImpPara_ Kerr--de Sitter}
\end{align}
Thus, the shadow cast by the two Kerr--de Sitter black holes (\ref{metric_ original}) and (\ref{metric}) are likely to be different. In the limit $\Lambda\to0$, both Equations~(\ref{CriImpPara}) and (\ref{CriImpPara_ Kerr--de Sitter}) go to the critical impact parameters of the Kerr black hole \citep{Chandrasekhar:1985kt}.

Before proceeding further, we note that $\eta_{crit}=0$ leads to planar circular orbits limited to the equatorial plane \citep{Chandrasekhar:1985kt,Teo:2020sey,Kumar:2020yem,Johnson:2019ljv}.
The allowed region for photon motion around the black hole (\ref{metric}) is given by $\mathcal{R}\geq 0$ and $\Theta(\theta)\geq 0$; considering $\eta_{crit}=0$ and $\Theta=0$, we obtain the photon orbit radii ($r_p^{\mp}$) of the new Kerr--de Sitter black holes and the photon region is given by
\begin{align}
	r_{p}^{-}\leq r_p\leq r_{p}^{+};\;
    r_{p}^{\mp}\equiv&\alpha\left[1+\sqrt{1-4a^2\Lambda}\cos\left(\frac{1}{3}\arccos\left(\frac{\mp\left|\beta\right|}{9M^2}\right)\right)\right] ,\nonumber\\
 \theta_{-}\leq \theta\leq \theta_{+};\;\;\theta_{\mp}\equiv&\arccos{\left[\mp \left(\frac{a^2-\eta_{crit} -\xi_{crit}^2+\sqrt{\left(a^2-\eta_{crit} -\xi_{crit}^2\right)^2+4 a^2 \eta_{crit} }}{2 a^2}\right)^\frac{1}{2}\right]}, \nonumber\\
 \text{with}\;\;\alpha=\frac{6M}{3+4a^2\Lambda}& \;\text{and}\;
    \beta=\frac{9 M^2-32 a^6 \Lambda ^2-48 a^4 \Lambda +18 a^2 \left(6 \Lambda  M^2-1\right)}{9 M^2 \left(4 a^2 \Lambda -1\right)\sqrt{1-4 a^2 \Lambda } }\ .
\end{align}
Here, $r_{p}^{\mp}$ are, respectively, the prograde and the retrograde photon radii around the new Kerr--de Sitter black holes (\ref{metric}) which are different from the Kerr--de Sitter (\ref{metric_ original}) \citep{Li:2020drn}. However, they go over to the photon sphere radii in the Kerr black hole ($\Lambda=0$) \citep{Kumar:2020ltt}:
\begin{equation}
    r_{p}^{k^{\mp}}\equiv2M\left[1+\cos\left({\frac{2}{3}\arccos\left(\mp\frac{|a|}{M}\right)}\right)\right],
\end{equation}
which are exactly same as that obtained by Teo \citep{Teo:2020sey}.
They fall in the range $M\leq r_{p}^{k^{-}}\leq 3M\leq r_{p}^{k^{+}}\leq 4M$ \citep{Kumar:2018ple} and further degenerate into a photon sphere of constant radius  $r_{p}^{k^{-}}=r_{p}^{k^{+}}=3M$  for the Schwarzschild black hole ($\Lambda=0, a=0$) \citep{Teo:2020sey,Kumar:2020hgm}. The photon region of the new Kerr--de Sitter black holes (\ref{metric}) is compared with the that of the Kerr--de Sitter black holes (\ref{metric_ original}) (cf. Figure~\ref{photonRegion_plot}). The event horizon of the former is also smaller than that of the latter (cf. Figure~\ref{photonRegion_plot}).

 The unstable photons form successive photon \textit{subrings} that asymptotically spiral and approach the boundary of  a dark silhouette called the shadow of the black hole \citep{Johnson:2019ljv}. In the present study, we shall investigate how the warped curvature of the new Kerr--de Sitter spacetime (\ref{metric}) manifests observably itself through the shadow geometry and calculate the deviation of its shadow observables from those of the  Kerr--de Sitter black hole observables. We assume a static observer at a finite location from the black hole ($r_o, \theta_o$)  such that  $r_{+}^{NK}<r_o<r_{c}^{NK}$; considering the domain of outer communication filled with light sources everywhere except between the observer and the black hole \citep{Grenzebach:2014fha}, we shall only focus on the unstable photon orbits in the exterior photon region ($r>r_{+}^{NK}$) \citep{Grenzebach:2014fha}.
In order to define the celestial coordinates along the shadow boundary, we construct the orthonormal basis at the observer position \citep{Grenzebach:2014fha,Kumar:2017tdw} as
\begin{align}
e_0=&\frac{(r^2+a^2)\partial_t+a\partial_{\phi}}{\sqrt{\Sigma\Delta}},\;\;e_1=\frac{1}{\sqrt{\Sigma}}\partial_{\theta},\nonumber\\
e_2=&-\frac{\partial_{\phi}+a\sin^2{\theta}\partial_t}{\sqrt{\Sigma}\sin\theta},\;\; e_3=-\sqrt{\frac{\Delta}{\Sigma}}\partial_r.
\end{align} 
By making a stereographic projection of the black hole shadow on the observer's celestial sky to the image plane, the shadow boundary is outlined by the celestial coordinates~\citep{Grenzebach:2014fha,Kumar:2017tdw}:
\begin{align}
\sin\chi(r_o,r_p)=&\frac{\xi_{crit}(r_p)-a}{\sqrt{(a-\xi_{crit}(r_p))^2+\eta_{crit}(r_p)}},\nonumber\\ 
\sin \Phi(r_o,r_p)=&\frac{\sqrt{\Delta[(a-\xi_{crit})^2+\eta_{crit}]}}{r^2+a^2-a\xi_{crit}},\label{Celestial1}
\end{align} 
which in the Cartesian plane become: \citep{Grenzebach:2014fha,Kumar:2017tdw},
\begin{align}
X(r_o,r_p)=&-2 \tan \left(\frac{\Phi(r_o,r_p)}{2}\right) \sin\left(\chi(r_o,r_p)\right),\nonumber\\
Y(r_o,r_p)=&-2 \tan \left(\frac{\Phi(r_o,r_p)}{2}\right) \cos\left(\chi(r_o,r_p)\right).\label{Celestial2}    
\end{align}
Using the set of all ($\xi_{crit}, \eta_{crit}$) obtained in Equation~(\ref{CriImpPara}) in Equation~(\ref{Celestial2}), we produce a parametric closed curve in the ($X-Y$) plane which limns the shadow of the black hole.
\begin{figure}
	\begin{adjustwidth}{-\extralength}{0cm}
\centering
		\begin{tabular}{c c}
			\hspace{-0.6cm}\includegraphics[scale=0.87]{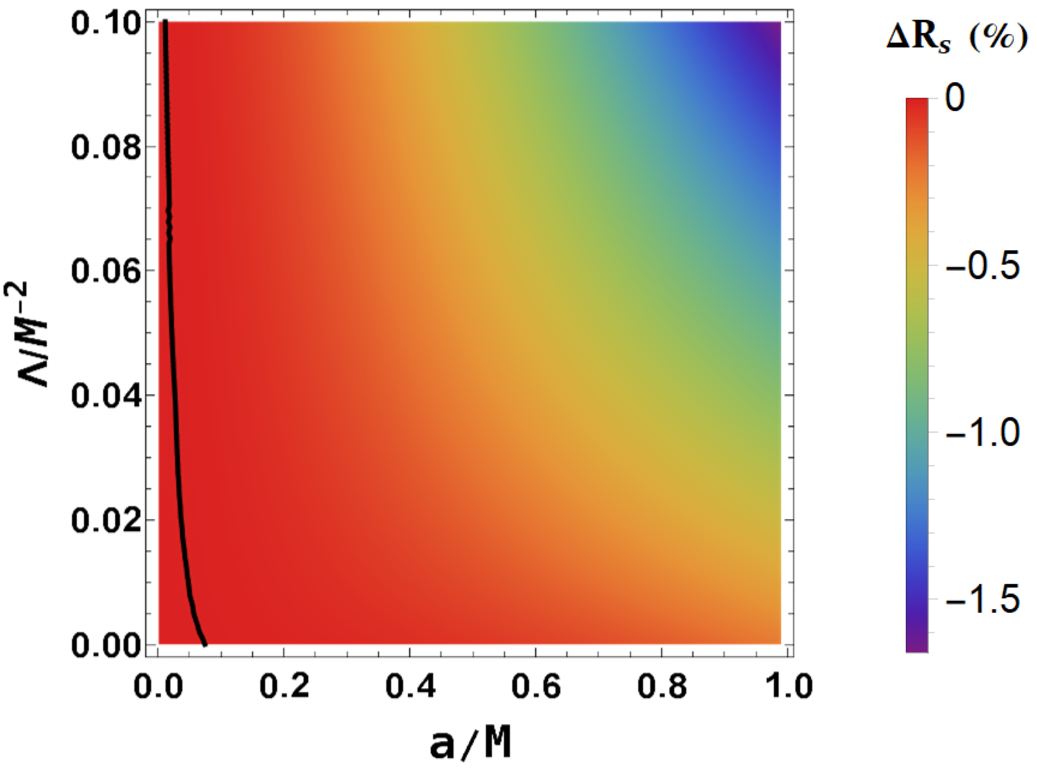}&
		\hspace{-0.6cm}\includegraphics[scale=0.87]{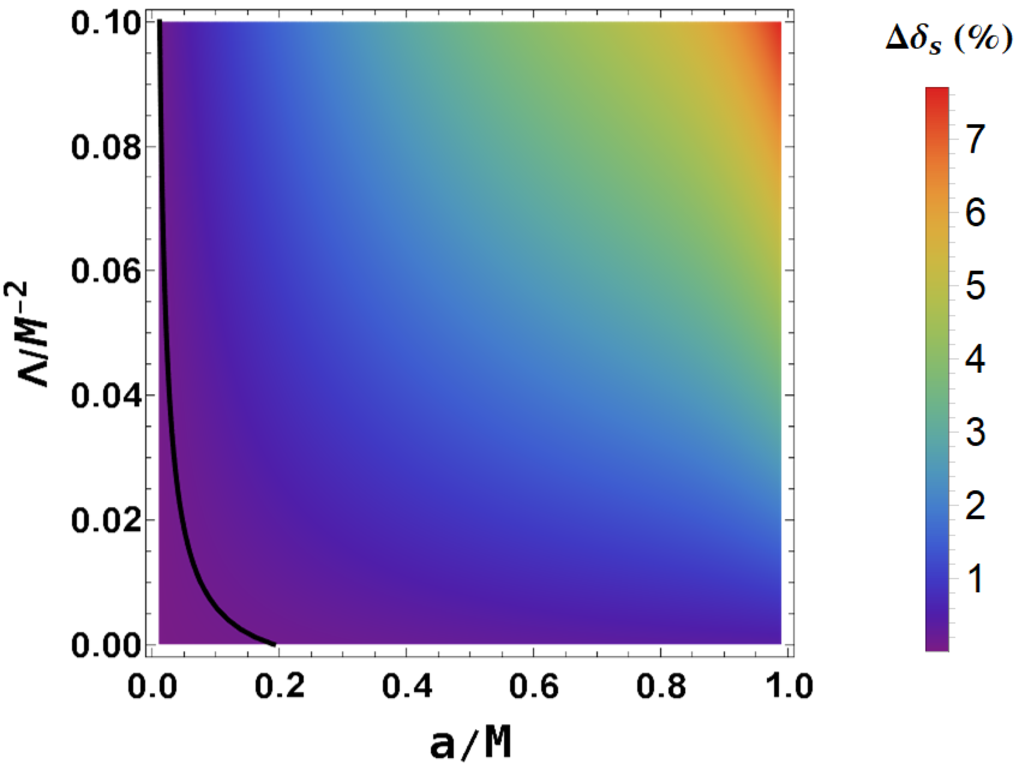}\\
			\hspace{-0.6cm}\includegraphics[scale=0.87]{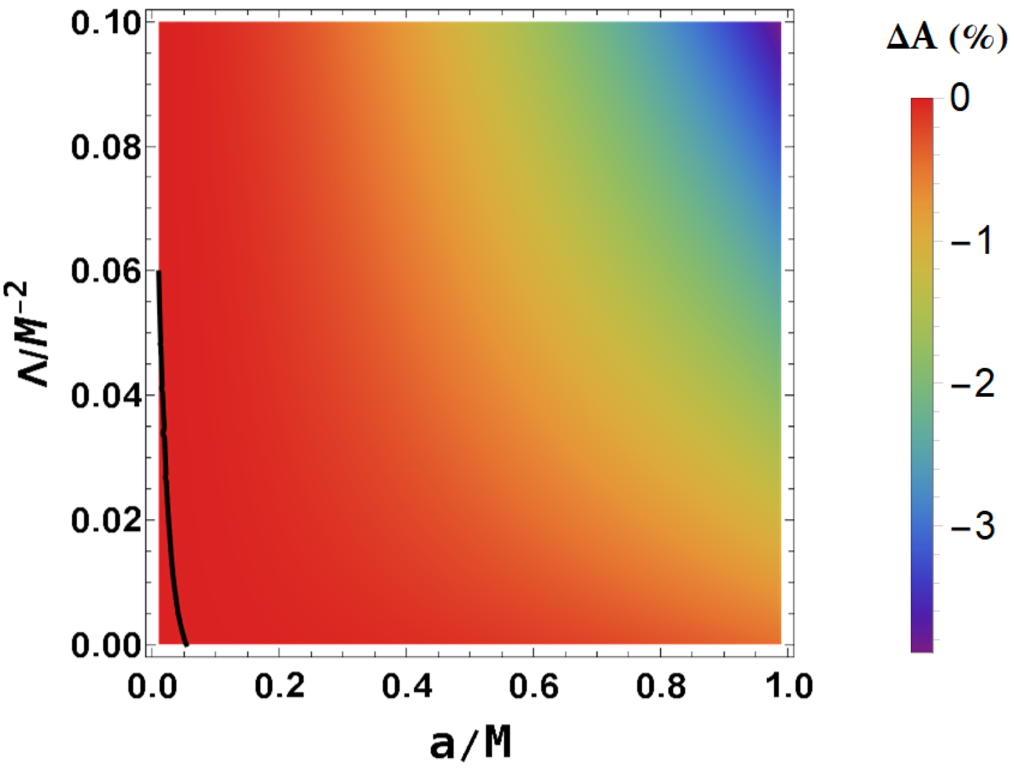}&
			\hspace{-0.6cm}\includegraphics[scale=0.87]{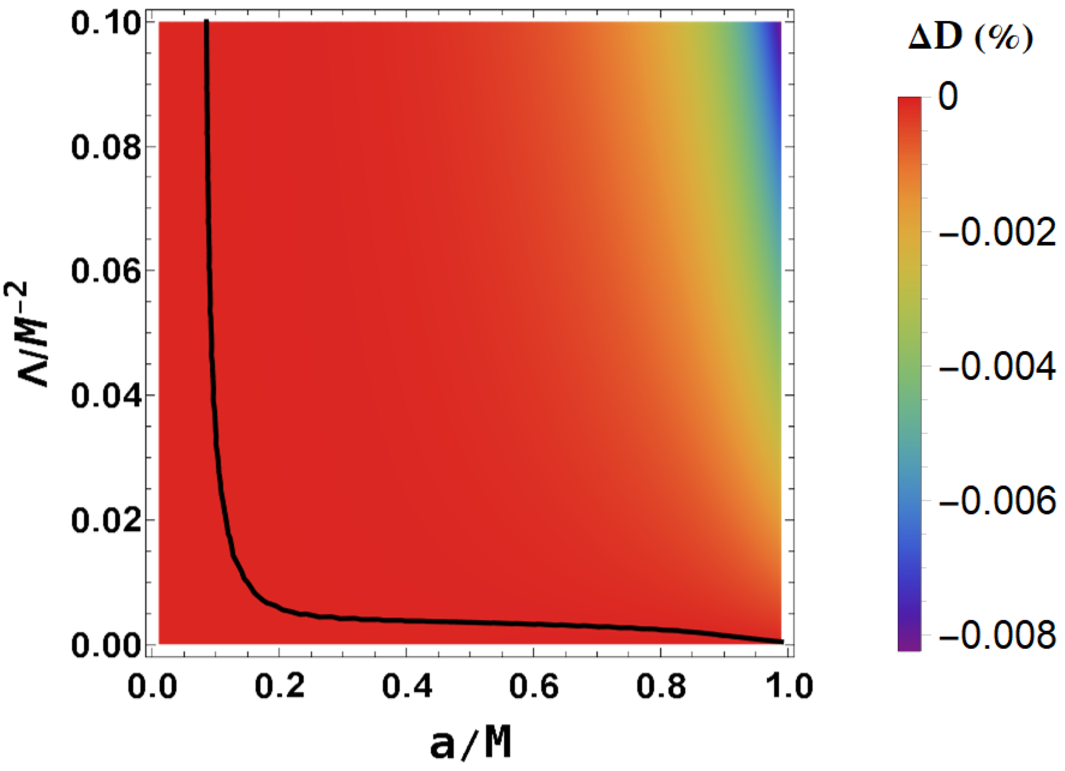}
		\end{tabular}
\end{adjustwidth}
\caption{Relative difference $\Delta \mathcal{O}$ ($\%$) between the shadow observables of the new Kerr--de Sitter black holes and the  Kerr--de Sitter black hole  at $(r_0, \theta_o) = (4M, 90^{\circ})$. In the regions bounded within the black solid curves and the axes, the observables of the new Kerr--de Sitter black holes  and the Kerr--de Sitter black hole are indistinguishable.}
	\label{ParameterComparison}
\end{figure}

For a static observer, the $\Lambda$ is found to have a decremental effect on the shadow size in both the Kottler black holes as well as the new Kerr--de Sitter black holes (cf.  Figure~\ref{shadow}) which is similar to the behaviour of the Kerr--de Sitter black hole shadows \citep{Li:2020drn}. Moreover, for fixed values of $a$ and $\Lambda$, the shadows are found to decrease in size as the static observer retreats from $r_{p}^{+}$ to $r_{c}^{NK}$ (cf.  Figure~\ref{shadow1}); when the observation distance varies from $r_p^+$ to $r_{+}^{NK}$, the shadows become increasingly deformed in the prograde region (cf.  Figure~\ref{shadow1}) and finally, at $r_o=r_{+}^{NK}$, the shadow disappears  and the observer sky becomes completely dark as in the Kerr--de Sitter case  \citep{Perlick:2018iye,Kumar:2017tdw}.
\section{Parameter Estimation and Relative Difference of Shadow Observables}\label{Sec4}
Estimation of the black hole parameters has been under constant endeavour \citep{Roelofs:2021wdi,Broderick:2021ohx,EventHorizonTelescope:2020eky,Feng:2017vba,Narayan:2007ks,Narayan:2005ie} and the shadows have been utilized to estimate the parameters in both GR \citep{Hioki:2009na,Tsukamoto:2014tja,Tsupko:2017rdo} as well as in modified theories of gravity (MoGs) \citep{Kumar:2018ple,Afrin:2021wlj,Ghosh:2020spb}. Interestingly, though the black hole spin and various other hairs have been estimated, the shadow has not been utilized for the estimation of the cosmological constant $\Lambda$ thus far. In the present study, we intend to estimate the $\Lambda$ in addition to the spin using shadow observables.
Einstein introduced the cosmological constant $\Lambda$
to balance the evolutionary models with repulsion to set a
steady-state but later abandoned as a blunder. One can introduce $\Lambda$ as the vacuum energy required to drive
inflation \citep{Akcay:2010vt}. The accelerating expansion may even be interpreted as the continuation of inflation, possibly at a slower rate than in the early universe. The cosmological constant $\Lambda$ has a dimension  $(length)^2$, and as per the classical general relativity, one has no preferred choice of what the length scale defined by $\Lambda$ could be.

Hioki and Maeda approximated the shadow silhouette to a reference circle coinciding with the top ($X_t$, $Y_t$), bottom ($X_b$, $Y_b$) and right ($X_r, 0$) extremes of the shadow; the shadow size and deviation from a perfect circle was thus characterized by two observables, the shadow radius ($R_s$) and distortion ($\delta_s$) \citep{Hioki:2009na} which are defined as
\begin{eqnarray}
    R_s=\frac{(X_t-X_r)^2+Y_{t}^2}{2|X_r-X_t|},\;\; \delta_s=\frac{|X_l-X^{'}_l|}{R_s},
\end{eqnarray}\label{Rs_delats}
where the shadow symmetry concerning the $x$ axis implies $(X_b,Y_b)=(X_t,-Y_t)$. Furthermore, the reference circle and leftmost edge of the shadow silhouette intersect with the $x$ axis at ($X^{'}_l$, 0) and ($X_l$, 0), respectively, \citep{Hioki:2009na}. However, the requirement of the shadow to be approximated by a circle is found to not always be feasible viz., for haphazard shadows in MoGs \citep{Schee:2008kz,Johannsen:2015qca,Tsukamoto:2014tja, Abdujabbarov:2015xqa,Younsi:2016azx, Tsupko:2017rdo} and also for noisy data \citep{Abdujabbarov:2015xqa,Kumar:2018ple}. Thus, more versatile observables---the shadow area ($A$) and oblateness ($D$)---characterizing a general haphazard shadow were introduced by Kumar and Ghosh \citep{Kumar:2018ple} which only require a symmetry concerning the $x$ axis and are defined as
\begin{align}
A=2\int_{r_p^{-}}^{r_p^+}\left( Y(r_p) \frac{dX(r_p)}{dr_p}\right)dr_p\;
,\;
D=\frac{X_r-X_l}{Y_t-Y_b}.\label{Area_Oblateness}
\end{align}

The shadow observables of the new Kerr--de Sitter black holes are found to be impacted by both the spin $a$ as well as the $\Lambda$ (cf.  Figure~\ref{Observables_plot}).
Taking a static observer at $r_o=4M$ observing at an inclination of $\theta_o=90^{\circ}$, we plot the contours of the two pairs of observables ($R_s$, $\delta_s$) and ($A/M^2, D$) (cf.  Figure~\ref{parameterEstimation_plot}) in the ($a/M$-$\Lambda /M^{-2}$) space---the unique contour intersection points thus obtained unambiguously determine the spin $a$ as well as the $\Lambda$ that we tabulate for few representative points (cf. Table~\ref{Table_parameter}).
Here, we propose a new method to estimate the value of $\Lambda$ using a black hole shadow, which may be useful in Cosmology. Figure \ref{parameterEstimation_plot} is representative, however, of a black hole with mass $M$, with which one can adjust the range of $\Lambda$ suitably. Considering the M87* as a new Kerr--de Sitter black hole, and using the mass $M=6.5\times10^9 M_{\odot}$ and distance $d=16.8$ Mpc as reported by the Event Horizon Telescope collaboration (EHT) \citep{Akiyama:2019bqs,Akiyama:2019cqa,Akiyama:2019eap}, we estimate the  value of the cosmological constant $\Lambda=1.046\times10^{-52} m^{-2}, 1.368\times10^{-52} m^{-2}$ obtained, respectively, for ($R_s$, $\delta_s$)= ($4.985\times 10^{13} m, 0.5$) and ($A$, $D$)= ($7.298\times 10^{27} m^2, 0.9812$), which agree with the present estimated value of $\Lambda=1.11\times10^{-52} m^{-2}$ \citep{Planck:2018vyg,Planck:2015fie,Stepanian:2021vvk}. Thus, our method is robust to estimate the cosmological constant $\Lambda$ as well as the spin $a$, provided one has a suitably chosen range for a given black hole.
Thus, we estimated, as a first, the cosmological constant using the black hole shadow which opens a new avenue to use the shadow for cosmological estimations.

Finally, to compare and contrast the new Kerr--de Sitter black holes from the original Kerr--de Sitter black hole, we plot the relative difference between the shadow observables ($\Delta \mathcal{O}$) of the two Kerr--de Sitter black holes.  In the ($a/M$-$\Lambda /M^{-2}$) space, the $\mathcal{O}^{NK}$ are identical to $\mathcal{O}$ over a finite parameter space (the region bounded by the black lines with the axes in  Figure~\ref{ParameterComparison}). An intersection of the four regions ($\mathcal{I}$) where the $\Delta \mathcal{O}=0$ would give a bound on the parameters $a$ and $\Lambda$ of the new Kerr--de Sitter black holes for which the shadow cast by them perfectly capture the shadow of Kerr--de Sitter black holes for any static observer at a given distance and observation angle. On the other hand, in the region $\{0\leq a/M\leq 1$, $0\leq \Lambda /M^{-2}\leq 0.1$\}$\backslash \mathcal{I}$, one would be able to distinguish the two shadows and consequently tell the two Kerr--de Sitter black holes (\ref{metric_ original}) and (\ref{metric}) apart (cf.  Figure~\ref{ParameterComparison}).

\section{Conclusions}\label{conclusions}
We explored the various spacetime properties in the strong-field regime of the new Kerr--de Sitter black holes and found the curvature's warping effect. The horizons significantly deviate from the original Kerr--de Sitter horizons. We obtained an expression for the photon sphere radii and found that they are different from those in the original Kerr--de Sitter. Consequently, they impact the photon region's structure in the new Kerr--de Sitter black holes and the black hole shadows. The new Kerr--de Sitter black hole, not asymptotically flat, has a cosmological horizon which prevents us from considering a static observer at an arbitrarily large distance and is compelling to restrict it within the region ($r_+,r_c$). Thus, our formalism allows for observers at any Boyer--Lindquist coordinates in the domain of outer communication. We derived an analytical formula for the new Kerr--de Sitter black hole shadows using a detailed analysis of the photon regions.
We present observables $A$ along with $D$ characterizing the new  Kerr--de Sitter shadow shape to estimate the cosmological constant $\Lambda$. Interestingly, we find a finite parameter space for ($\Lambda$, $a$) where the two observables $A$ and $D$ of the new Kerr--de Sitter black holes and the Kerr--de Sitter black hole are indistinguishable.

Finally, our results, in the limit $\Lambda \to 0$, go over to those of the Kerr black holes.

\textbf{Note added in Proof}: After this work was submitted to the Journal, we learned of a similar work by E. Omwoyo et al. \citep{Omwoyo:2021uah}, which appeared in arXiv a couple of days after submission.




\authorcontributions{Conceptualization, M.A. and S.G.G.; data curation, M.A. and S.G.G.; formal analysis, M.A. and S.G.G.; funding acquisition, M.A. and S.G.G.; investigation, M.A. and S.G.G.; methodology, M.A. and S.G.G.; project administration, S.G.G.; resources, M.A. and S.G.G.; software, M.A. and S.G.G.; supervision, S.G.G.; validation, M.A. and S.G.G.; visualization, M.A. and S.G.G.; writing---original draft, M.A. and S.G.G.; writing---review and editing, M.A. and S.G.G. All authors have read and agreed to the published version of the manuscript.
}

\funding{M.A. is supported by the DST-INSPIRE fellowship from the Department of Science and Technology, Govt. of India.
}

\institutionalreview{Not applicable.}

\informedconsent{Not applicable.}

\dataavailability{We have not generated any original data in the due course of this
study, nor has any third-party data been analysed in this article.}

\conflictsofinterest{The authors declare no conflict of interest.}


\abbreviations{Abbreviations}{
The following abbreviations are used in this manuscript:\\

\noindent
\begin{tabular}{@{}ll}
GR & General Relativity\\
MoGs & Modified Theories of Gravity
\end{tabular}}
\begin{adjustwidth}{-\extralength}{0cm}

\reftitle{References}
\bibliography{sample63.bib}

\end{adjustwidth}
\end{document}